\documentclass[thmsb]{article}
\usepackage{amssymb}

\usepackage{sw20aip}

\input{tcilatex}
\begin{document}

\author{H. Bergeron\\
\\
LURE Bat. 209D\\
Centre Universitaire Paris-Sud -BP34- 91898 Orsay Cedex\\
e-mail: bergeron@lure.u-psud.fr}
\title{From Classical to Quantum Mechanics:\\
''How to translate physical ideas into mathematical language''\\
}
\date{Submitted to Journal of Mathematical Physics; Dec. 20, 2000}
\maketitle

\begin{abstract}
In this paper, we investigate the connection between Classical and Quantum
Mechanics by dividing Quantum Theory in two parts:

- General Quantum Axiomatics (a system is described by a state in a Hilbert
space, observables are self-adjoints operators and so on...)

- Quantum Mechanics properly that specifies the Hilbert space as $L^{2}(%
\mathbf{R}^{n})$; the Heisenberg rule $[\mathbf{p}_{i},\mathbf{q}
_{j}]=-i\hslash $ $\delta _{ij}$ with $\mathbf{p}=-i\hslash \nabla $, the
free Hamiltonian $\mathbf{H}=-\hslash ^{2}\Delta /2m$ and so on.

We show that General Quantum Axiomatics (up to a supplementary ''axiom of
classicity'') can be used as a non-standard mathematical ground to formulate
all the ideas and equations of ordinary Classical Statistical Mechanics. So
the question of a ''true quantization'' with ''$\hslash $'' must be seen as
an independent problem not directly related with quantum formalism.

The idea of using Hilbert space techniques in Classical Mechanics is not
new; it was introduced by Koopman\cite{Koopman} and has been extensively
employed in Statistical Mechanics\cite{hilbertclass}$^{,}$\cite{schonberg}$%
^{,}$\cite{prigogine1} . But in this article, we don't look at this
technique as being only a mathematical trick: we consider that the Hilbert
space formalism is really a mathematical and physical ground to rebuild
Classical Mechanics.

At this stage, we show that this non-standard formulation of Classical
Mechanics exhibits a new kind of operation that has no classical
counterpart: this operation is related to the ''quantization process'', and
we show why quantization physically depends on group theory (Galileo group).
This analytical procedure of quantization replaces the ''correspondence
principle'' (or canonical quantization) and allows to map Classical
Mechanics into Quantum Mechanics, giving all operators of Quantum Dynamics
and Schr\"{o}dinger equation.

The great advantage of this point of view is that now, quantization is based
on concrete physical arguments and not derived from some ''pure algebraic
rule'' (we exhibit also some limit of the correspondence principle).

Moreover spins for particles are naturally generated, including an
approximation of their interaction with magnetic fields. We find also that
this approach gives a natural semi-classical formalism: some exact dynamical
quantum results are obtained only using classical-like formula.

So this procedure has the nice property of enlightening in a more
comprehensible way both logical and analytical connection between classical
and quantum pictures.
\end{abstract}

\section{Introduction}

From the beginning of Quantum Mechanics, different methods have been
developed to link classical and quantum formalisms. One of the most famous
is the Wigner-Weyl transformation that allows to recover the semi-classical
limit of Quantum Mechanics in phase space\cite{Weyl} . This operation
associates to each quantum operator $\mathbf{a}$ a function in phase space $%
a(p,q)$ in such a way that at the lowest order in $\hbar $ (zero order), the
quantum evolution of $a(p,q)$ reduces to the classical one.\newline
In this point of view, quantization is mathematically a deformation of the
abelian algebra of functions in phase-space into a non-commutative algebra
using the so-called $*_{h}$-product that replaces standard multiplication%
\cite{deformalgebra} ($a(p,q)*_{h}b(p,q)$ corresponds to the operator
product $\mathbf{a}.\mathbf{b}$).\newline
Although the Wigner-Weyl transformation is algebraically very powerful, it
is not completely satisfactory on a physical point of view, because all
quantum densities $\mathbf{D}$ are not mapped into true probability densities%
\cite{wignerpositive} $\rho (p,q)$ (positive functions). Nevertheless, this
formalism is very useful, and not only in non-relativistic mechanics, since
it can be extended to Special Relativity\cite{mourad} .

The Wigner-Weyl transformation is not the unique way to obtain
semi-classical formula. Such kind of formula also occurs using a coherent
state splitting\cite{Klauder} . Moreover we have already proved that we can
define Hamiltonian-dependent coherent states that allow to obtain exact
quantum results with semi-classical formula\cite{berg} , and this approach
preserves positivity (a quantum density $\mathbf{D}$ is associated with a
true classical probability density $\rho (p,q)$). This formalism with
coherent states can be also extended to relativity\cite{prugo} .

All these procedures have the nice property of giving an analytical
connection between classical and quantum formalisms, overcoming the apparent
discontinuity between classical and quantum pictures. Nevertheless, we have
always to assume first quantum operators to obtain semi-classical results,
while at the same time, quantum operators (and specially quantum
Hamiltonian) are themselves derived from the correspondence principle. So
quantum operators are in fact deduced from classical ones by some ''pure
algebraic rule''. Then, at a logical level, relationship between Classical
and Quantum Mechanics remains not so clear, since we need both of them at
the same time, although Quantum Mechanics is assumed in principle to be the
''only true''.

This shows that logical relationship between Quantum Mechanics and Classical
Mechanics is more complicated than, for example, relationship between
Classical Mechanics and Special Relativity, while Quantum Mechanics needs
only a new constant ''$\hslash $'', exactly as Special Relativity needs the
new constant ''$c$''. We mean that Classical and Relativistic Mechanics are
formulated on the same mathematical ground (basic mathematical objects) in
such a way that taking the limit $c\rightarrow \infty $ into any
relativistic formula, we obtain the standard classical result. But this
cannot be done directly in the same way ($\hslash \rightarrow 0$) in the
case of Quantum Mechanics because the Hilbert space formalism is not the
mathematical formalism of Classical Mechanics.\newline
For the same reason, we don't need in Special Relativity a general procedure
to lift up a classical quantity into the relativistic frame, while in
Quantum Mechanics, we need the correspondence principle.

Of course, different studies have been led, not on a purely analytical
level, but on a logical level, to find an a priori justification of the
quantum picture and these attempts go back to von Neumann\cite{Neumann} .%
\newline
More recently, in the sixties, Mackey\cite{Mackey1}$^{,}$\cite{Mackey2} has
presented a general mathematical description of a statistical system based
on an abstract structure called ''orthocomplemented lattice'' that unifies
classical and quantum statistics: adding an ad hoc postulate to the basic
lattice picture allows to recover either classical formalism or quantum one.%
\newline
This shows that the mathematical representations of classical and quantum
theories share some general common structure, but this does not explain the
special realization leading to Quantum Mechanics, namely the choice of the
Hilbert space as $L^{2}(\mathbf{R}^{n})$ with $\mathbf{p}=-i\hslash \nabla $
and the free Hamiltonian $\mathbf{H}=-\hslash ^{2}\Delta /2m$. In fact,
different studies show that these special realizations are connected to
group theory, namely Euclidean invariance and Galilean invariance\cite
{Mackey2}$^{,}$\cite{Groups} .\newline
To conclude this short overview, let us say that a lot of work has been done
on these ''first level'' foundations of Quantum Mechanics and we can only
give a non-exhaustive list of references\cite{qaxioms} .

Now, let us situate our article and its goals.\newline
As noticed previously, the difficulty to connect Classical and Quantum
Mechanics on both logical and analytical levels is essentially due to the
great difference of mathematical language that confuses physical intuition.
Our main goal is to develop some precise mathematical ground in which
Classical Mechanics and Quantum Mechanics can be expressed using the same
tools, in such a way that physical ideas and logical arguments recover the
first place in comparison with ''obscure algebraic rules''. Moreover we want
to obtain a precise status (classical or quantum) for each ingredient of
Quantum Mechanics.\newline
This question of the classical or purely quantum nature of the ingredients
of Quantum Mechanics is legitimate by at least two arguments:\newline
- First, most of quantum operators are deduced from classical quantities
(correspondence principle).\newline
- Secondly, taking the limit value $\hslash \rightarrow 0$, Quantum
Mechanics must degenerate into Classical Mechanics. Unfortunetaly, this
limit only says that $\mathbf{p}$ and $\mathbf{q}$ are now commuting
operators: the general quantum formalism with wave functions, operators and
so on, remains unchanged.

The only possible explanation of this last result, is that quantum formalism
with commuting operators, describes the ideas of Classical Statistical
Mechanics as well as the usual formalism in phase space. This means that we
have in fact two possible mathematical formalisms to represent the same
physical ideas of Classical Statistical Mechanics. Then it must be possible
to formulate the ideas and the equations of standard Classical Statistical
Mechanics only using the general quantum formalism (axiomatics) and then the
question of the true quantization with $\hslash $ must be seen as an
independant problem not directly related with the formalism.

In this article, we present such a ''non-standard formulation'' of Classical
Statistical Mechanics and we show how (and why) it is equivalent to usual
one. Of course this new picture is not a true quantum theory, since it
represents always classical ideas; but it works with general quantum
formalism. As indicated in the abstract, the use of Hilbert space techniques
into classical Mechanics is not new and goes back to Koopman\cite{Koopman} .
Even the concept of ''classical amplitude'' is not new and has been
introduced earlier by Sch\"{o}nberg\cite{schonberg} . But in this article we
don't look at these techniques with the same point of view: we consider the
Hilbert space formalism as a true logical and analytical representation of
Classical Mechanics, and not only as a mathematical trick.

Then, we show that this new formalism exhibits a new kind of operation that
has no classical counterpart: this operation is related to the
''quantization process'', and we show why quantization physically depends on
group theory (Galileo group).\newline
This procedure replaces the correspondence principle and allows to recover
all operators of Quantum Mechanics and Schr\"{o}dinger equation. The great
advantage of this point of view is that now, quantization is explicitly
based on concrete physical arguments and not derived from some ''pure
algebraic rule''. Moreover spins for particles are naturally generated,
including an approximation of their interaction with magnetic fields. We
find also that this approach gives a natural semi-classical formalism: some
exact quantum results are obtained with classical-like formula.

To conclude this introduction, let us specify two points.\newline
First, we don't situate our article in the frame of the question: why
macroscopic world appears essentially classical, while microscopic world is
quantum? It is well-known that this effect is due to decoherence, for
details see this non-exhaustive list of references\cite{decoherence} . Our
article is not any more in the field of ''consistent interpretations of
Quantum Mechanics'' or ''consistent histories'' that looks for the logical
problems raised by Quantum Mechanics (Omn\`{e}s\cite{Omnes} ,
Gell-Mann-Hartle \cite{GellMann} , Griffiths\cite{Griffiths} ), even if our
article contains consequences on the interpretation of quantization.

Secondly, there exists one important aspect of our point of view that is not
developed in this article: this concerns our previous remark on the
equivalence of quantum formalism with commuting operators and usual
classical formalism. If this equivalence is true, this means that any one of
these formalisms can be (logically and physically) rebuilt from the other
one, and more specially that the full general quantum frame (axiomatics) can
be found (as a change of mathematical language) only starting with classical
formalism. Of course, on a logical level, it should be more enlightening to
recover first general quantum formalism, before any further development. But
this should be too long for a single article. So, this special point will be
published later in an article devoted to this question.\newpage

The article is organized as follows:\textbf{\newline
I Introduction\newline
II Classical Mechanics in Phase Space}

\textbf{A }Phase Space Structure and Poisson Brackets

\textbf{B} Classical Dynamics

\textbf{C} Physical Symmetries in Phase Space\textbf{\newline
III New Framework for Classical Mechanics}

\textbf{A} The Basic Elements

\textbf{B} Equivalence of the New Frame with Standard Formalism

\textbf{C} Conclusion\newline
\textbf{IV Representation of Poisson Brackets in the New Framework}\newline
\textbf{V Classical Dynamics in the New Framework}

\textbf{A} New Formulation of Equations of Motion

\textbf{B} Classical Dynamics

\textbf{VI The Quantization Process}

\textbf{A} The problem of Unclassical Observables: what is quantization?

\textbf{B} Consequences of a Quantization

\textbf{C} Our Arguments to choose a Quantization\textbf{\newline
VII Representations of Galileo Group}

\textbf{A} The Classical Representation

\textbf{B} First Consequences of a Unit of Action

\textbf{C} New Representation of Galileo Group

\textbf{D} Angular Momentum and Spin

\textbf{E} Discrete Symmetries\textbf{\newline
VIII} \textbf{Irreducible Representation of Galileo Group}

\textbf{A} Subspaces of irreducible representations of $P_{*}$ and $Q_{*}$

\textbf{B} Irreducible representation of rotations

\textbf{C} Conclusion

\textbf{D} Action of Discrete Symmetries on an Irreducible Subspace\textbf{%
\newline
IX} \textbf{Axiom of Quantization\newline
X The Basic Quantized Observables}

\textbf{A} Quantum Operators of Position and Momentum

\textbf{B} Quantum Angular Momentum and Spin

\textbf{C} Semi-Classical States and Semi-Classical Events

\textbf{D} Semi-Classical States and Quantum Observables

\textbf{E} Quantized Observables and Statistics\textbf{\newline
XI Quantum Dynamics}

\textbf{A} Quantum Operator of Evolution: Schr\"{o}dinger Equation

\textbf{B} Quantum Hamiltonian\textbf{\newline
XII Conclusion}

\newpage

\section{Classical Mechanics in Phase Space (one particle)}

\subsection{Phase Space Structure and Poisson Brackets}

We only present in this paragraph the main features that we need, for more
details, see\cite{mecaclassique} .

In Classical Mechanics, Phase Space represents Configuration Space and it is
the set of pairs $(\overrightarrow{p},\overrightarrow{q})$ of momentum and
position, where $(\overrightarrow{p},\overrightarrow{q})$ represents the
(pure) state of the system. Any physical observable of the system is given
by a function $f(\overrightarrow{p},\overrightarrow{q})$ on Phase Space:
this is a consequence of the physical hypothesis specifying that the data of
momentum and position determines completely the system.\newline
Now, from a mathematical point of view, if we call $M$ the $\mathbf{R}^{3}$
space manifold, Phase Space is the cotangent bundle $TM^{*}$. It possesses a
natural geometry (namely a symplectic geometry) that allows to define the
Poisson Brackets (PB), $\left\{ f,g\right\} $ of two functions by: 
\begin{equation}
\left\{ f,g\right\} =\overrightarrow{\nabla _{p}}f.\overrightarrow{\nabla
_{q}}g-\overrightarrow{\nabla _{p}}g.\overrightarrow{\nabla _{q}}f
\label{PoissonBr}
\end{equation}

PB is the basic algebraic component of Classical Mechanics and the physical
content of PBs can be related to group theory.

Namely, any classical observable $f$ can be seen as the generator of a
one-parameter group of transformations acting on observables. Any $g$ is
transformed into $g_{\alpha }$ through: 
\begin{equation}
\frac{\partial g_{\alpha }}{\partial \alpha }=\left\{ f,g_{\alpha }\right\}
\label{PBevol}
\end{equation}

Equivalently, we have trajectories of states $(\overrightarrow{p_{\alpha }},%
\overrightarrow{q_{\alpha }})$ generated by $f$ following the equations: 
\begin{equation}
\left\{ 
\begin{array}{l}
\frac{\partial }{\partial \alpha }\overrightarrow{q_{\alpha }}=%
\overrightarrow{\nabla _{p}}f(\overrightarrow{q_{\alpha }},\overrightarrow{
p_{\alpha }}) \\ 
\frac{\partial }{\partial \alpha }\overrightarrow{p_{\alpha }}=-%
\overrightarrow{\nabla _{q}}f(\overrightarrow{q_{\alpha }},\overrightarrow{
p_{\alpha }})
\end{array}
\right.  \label{Generalpqevol}
\end{equation}

Then, on a mathematical level, all classical observables play the same role
and this role is summarized by the algebra due to Poisson Brackets.

\subsection{Classical Dynamics}

Dynamics on Phase Space is defined as the one-parameter group induced by a
special observable $H$ called the Hamiltonian (in general $H$ can be time
dependent).

\subsubsection{Equations of Motion for a Pure State}

Following equations (\ref{Generalpqevol}), the evolution with time of a
state $(\overrightarrow{p},\overrightarrow{q})$ is given by: 
\begin{equation}
\left\{ 
\begin{array}{l}
\frac{d}{dt}\overrightarrow{q}=\overrightarrow{\nabla _{p}}H(\overrightarrow{
q},\overrightarrow{p},t) \\ 
\frac{d}{dt}\overrightarrow{p}=-\overrightarrow{\nabla _{q}}H(%
\overrightarrow{q},\overrightarrow{p},t)
\end{array}
\right.  \label{hamilequs}
\end{equation}

The general expression of $H$ is obtained when we consider a particle of
mass $M$ and charge $e$ in interaction with fields: 
\begin{equation}
H=\frac{1}{2M}\left( \overrightarrow{p}-e\overrightarrow{A}(\overrightarrow{q%
},t)\right) ^{2}+V(\overrightarrow{q},t)
\end{equation}

The potentiel energy $V(\overrightarrow{q},t)$ contains the electrostatic
term and possible other interactions.

\subsubsection{Dynamics and Statistics}

These equations (\ref{hamilequs}) correspond to the ideal case of a particle
perfectly localized in Phase Space and we can represent this situation by
the probability density $\rho (\overrightarrow{p},\overrightarrow{q}
,t)=\delta (\overrightarrow{p}-\overrightarrow{p_{0}}(t))\delta (%
\overrightarrow{q}-\overrightarrow{q_{0}}(t))$. Now, if we build a general
density $\rho $ as superpositions of ''$\delta $'' as $\rho
=\sum_{i}p_{i}\delta _{\overrightarrow{p_{i}}(t),\overrightarrow{q_{i}}(t)}$
, we find that $\rho $ verifies Liouville equation: 
\begin{equation}
\frac{\partial \rho }{\partial t}=-\left\{ H,\rho \right\}  \label{Liouville}
\end{equation}

So we say classically that (\ref{Liouville}) describes the evolution of any
probability density $\rho $.

Now, starting from a density $\rho $ that verifies (\ref{Liouville}), we can
look at the evolution of the expectation value $<f>_{t}$ of an observable $f(%
\overrightarrow{p},\overrightarrow{q},t)$ defined as: 
\begin{equation}
<f(\overrightarrow{p},\overrightarrow{q},t)>_{t}\text{ }=\int d^{3}%
\overrightarrow{p}d^{3}\overrightarrow{q}\rho (\overrightarrow{q},%
\overrightarrow{p},t)\text{ }f(\overrightarrow{p},\overrightarrow{q},t)
\end{equation}

After a few algebra, we find: 
\begin{equation}
\frac{d}{dt}<f>_{t}\text{ }=\text{ }<\frac{\partial f}{\partial t}
>_{t}+<\left\{ H,f\right\} >_{t}  \label{generalmeanequ}
\end{equation}

Applied to the special case of the two fundamental observables $%
\overrightarrow{p}$ and $\overrightarrow{q}$, the equation (\ref
{generalmeanequ}) gives: 
\begin{equation}
\left\{ 
\begin{array}{l}
\frac{d}{dt}<\overrightarrow{q}>_{t}\text{ }=\text{ }<\overrightarrow{\nabla
_{p}}H>_{t} \\ 
\frac{d}{dt}<\overrightarrow{p}>_{t}\text{ }=-<\overrightarrow{\nabla _{q}}
H>_{t}
\end{array}
\right.  \label{meanhamilequs}
\end{equation}

\subsubsection{Strong and Weak Dynamical Equations}

In this paragraph we want to point out some important remarks on the
dynamical equations in the frame of Statistical Mechanics.\newline
Let assume that the evolution of the probability density $\rho (%
\overrightarrow{p},\overrightarrow{q},t)$ is unknown, and that we only know
the evolution of expectation values of $\overrightarrow{p}$ and $%
\overrightarrow{q}$ through the equations (\ref{meanhamilequs}). We want to
see if it is possible to rebuild the equations (\ref{hamilequs}) on pure
states and to find Liouville equation, only starting with (\ref
{meanhamilequs}).

Basically, if we assume that ''$\delta $'' densities are allowed such as $%
\rho (\overrightarrow{p},\overrightarrow{q},t)=\delta (\overrightarrow{p}-%
\overrightarrow{p_{0}}(t))\delta (\overrightarrow{q}-\overrightarrow{q_{0}}
(t))$, we find that the equations (\ref{meanhamilequs}) on expectation
values imply that the trajectory $\left( \overrightarrow{p_{0}}(t),%
\overrightarrow{q_{0}}(t)\right) $ follows the hamiltonian equations (\ref
{hamilequs}), then we deduce in the same way that a general density must
verify Liouville equation.

But if ''$\delta $'' densities are not allowed, we cannot directly deduce
the evolution of states only starting with expectation value equations, and
then we cannot recover Liouville equation. This means that we can find other
possible laws of evolution for the density $\rho (\overrightarrow{q},%
\overrightarrow{p},t)$ that are compatible with the equations (\ref
{meanhamilequs}) on expectation values.

So we call equations (\ref{hamilequs}) on states \emph{''Strong Dynamical
Equations''} and equations (\ref{meanhamilequs}) on expectation values \emph{%
\ ''Weak Dynamical Equations''}.

Now, since any real result of a physical experiment contains always some
uncertainty, the true mathematical representation of the physical result is
a expectation value associated with some probability density. Then weak
dynamical equations (\ref{meanhamilequs}) are a better mathematical
description of our physical knowledge about classical dynamics. When we say
that strong dynamical equations (\ref{hamilequs}) are realized, in fact we
extrapolate our real knowledge, by assuming that we can use ''$\delta $''
densities. Of course this procedure is natural, but not logically necessary.
We will use this remark in paragraph V-A.

\subsection{Physical Symmetries in Phase Space}

We will see later in section VI that our quantization principle is based on
the representation of symmetries. Symmetries are defined as the set of
physical transformations allowing to look at the same system from different
but equivalent frames. Then let us specify in this part the symmetries
involved in Phase Space.

At first sight, the Poisson Brackets defined previously seem to prove that
each observable defines a generator of such a symmetry, and then the set of
symmetries is generated by the full set of classical observables.\newline
But this is only true on a mathematical level: it is false on a physical
point of view, because any of these transformations cannot be realized in
practice as a real change of frame. The only real continuous transformations
that can be realized are space translations, space rotations and galilean
transformations. They constitute the Galileo group. We can also add two
discrete transformations: parity and time reversal.\newline
So the true set of classical symmetries can be divided as follows:\newline
- The space transformations: translations, rotations and parity\newline
- The kinematical transformations: Galileo boosts.\newline
- Time reversal.

\section{New Framework for Classical Mechanics}

As indicated in the introduction, our point of view is to use standard
quantum formalism applied to the case of Classical Mechanics. Of course, it
could be more enlightening to show before that all the following axioms can
be directly generated by a change of mathematical language, only starting
with usual formalism of Classical Mechanics. But for reasons of concision,
we differ this development to a future article.\newline
So we introduce the orthonormal basis $\left\{ |\overrightarrow{p},%
\overrightarrow{q}>\right\} $ that diagonalizes at the same time the
operators $\overrightarrow{\mathbf{p}}$ and $\overrightarrow{\mathbf{q}}$ ($[%
\overrightarrow{\mathbf{p}},\overrightarrow{\mathbf{q}}]=0$). Now we list
the basic axioms\cite{standardqaxioms} (since we don't try to obtain a
minimal list, some axioms can be redundant).

\subsection{The Basic Elements}

\subsubsection{Primary Axioms}

\begin{itemize}
\item  The mathematical frame is the Hilbert space $\mathcal{H}=L^{2}(%
\mathbf{R}^{6})$ with the ''continuous orthogonal basis'' $\left\{ |%
\overrightarrow{p},\overrightarrow{q}>\right\} $:\textbf{\newline
}$<\overrightarrow{p},\overrightarrow{q}|\overrightarrow{p^{\prime }},%
\overrightarrow{q^{\prime }}>=$ $\delta (\overrightarrow{p}-\overrightarrow{%
p^{\prime }})\delta (\overrightarrow{q}-\overrightarrow{q^{\prime }})$.

\item  A particle is represented at each time by a normalized vector $\phi
_{t}$ in $\mathcal{H}$ called the state of the system.

\item  An observable $\mathbf{F}$ is a self-adjoint operator on $\mathcal{H}$
and the possible values of this observable are the eigen values of $\mathbf{F%
}$.
\end{itemize}

\emph{Remarks}\newline
Orthogonal projectors are special observables with eigen values $0$ and $1$
, so orthogonal projectors are logical observables.\newline
Moreover, by the spectral theorem\cite{spectraltheo} , an observable $%
\mathbf{F}$ is associated with a family of orthogonal projectors $\left\{ 
\mathbf{P}(A)\right\} _{A\in \mathcal{B}(\mathbf{R})}$ where $\mathcal{B}(%
\mathbf{R})$ is the family of Borel set in $\mathbf{R}$. The mapping $%
\mathbf{P}$ is called the Projection Valued Measure (pvm) associated with $%
\mathbf{F}$ and $\mathbf{F=}\int \lambda d\mathbf{P}_{\lambda }$.

\subsubsection{Statistical Axioms}

\begin{itemize}
\item  A statistical situation is described by a density operator $\mathbf{D}
$ (a positive trace class operator with $Tr\mathbf{D}=1$), in particular a
system in the state $\phi $ is statistically represented by the density $%
\mathbf{D}=|\phi ><\phi |$.

\item  The expectation value of an observable $\mathbf{F}$ is given by $<%
\mathbf{F}>=Tr(\mathbf{D.F})$, the standard error $\Delta F$ is given by $%
\Delta F^{2}=<\mathbf{F}^{2}>-<\mathbf{F}>^{2}$.
\end{itemize}

\subsubsection{Collapse Axiom}

\begin{itemize}
\item  Let a system in a situation described by the density $\mathbf{D}$ and
consider a measure of the observable $\mathbf{F}$ associated with its
projection valued measure $\mathbf{P}$. Now, assume that a measure of the
observable specifies that the numerical outcomes $f$ are in the range $f\in A
$, where $A$ is some interval. Then after the experiment, the system is
described by the new density operator $\mathbf{D}^{\prime }$ with: 
\begin{equation}
\mathbf{D}^{\prime }=\frac{1}{Tr(\mathbf{D}\text{ }\mathbf{P}(A))}\mathbf{P}%
(A)\text{ }\mathbf{D}\text{ }\mathbf{P}(A)
\end{equation}
More specially, if the system is initially in the state $|\phi >$, after the
experiment the system is in the state $|\phi ^{\prime }>$: 
\begin{equation}
|\phi ^{\prime }>=\frac{1}{\sqrt{<\phi |\mathbf{P}(A)|\phi >}}\mathbf{P}%
(A)|\phi >
\end{equation}
This means that after the experiment, the system is in a state that belongs
to the subspace of $\mathcal{H}$ associated with the orthogonal projector $%
\mathbf{P}(A)$.
\end{itemize}

\subsubsection{Axiom of Evolution}

\begin{itemize}
\item  The evolution of a state $\phi _{t}$ is given by a unitary operator $%
\mathbf{U}_{t_{1},t_{2}}$ such that $\phi _{t}=\mathbf{U}_{t,t_{0}}(\phi
_{t_{0}})$.\newline
\emph{Remark:}\newline
This last statement is not really a true axiom, since it can be logically
deduced from the other ones \cite{evolWigner} $^{,}$ \cite{evolSimon}
.\bigskip 
\end{itemize}

Now, to specify the situation of Classical Mechanics, we must add some new
axiom of ''classicity''.

\subsubsection{Axiom of Classicity}

\begin{itemize}
\item  The true physical observables are always diagonal in the $|%
\overrightarrow{p},\overrightarrow{q}>$ basis, and then a classical
observable $\mathbf{f}$ can be written: 
\begin{equation}
\mathbf{f}=\int d^{3}\overrightarrow{p}d^{3}\overrightarrow{q}\text{ }f(%
\overrightarrow{p},\overrightarrow{q})\text{ }|\overrightarrow{p},%
\overrightarrow{q}><\overrightarrow{p},\overrightarrow{q}|
\end{equation}
Or $\mathbf{f}=f(\overrightarrow{\mathbf{p}},\overrightarrow{\mathbf{q}})$
as a function of operators.\newline
This axiom is equivalent to the classical hypothesis specifying that all
observables are functions of $\overrightarrow{p}$ and $\overrightarrow{q}$.%
\newline
\emph{Remark:}\newline
Density operators $\mathbf{D}$ are not considered as observables and then,
they don't need to be diagonal in the $\left\{ |\overrightarrow{p},%
\overrightarrow{q}>\right\} $ basis (in fact this is impossible for a trace
class operator).\bigskip 
\end{itemize}

Now, we claim that this new mathematical frame constitutes a complete
alternative formulation of Classical Statistical Mechanics.

\subsection{Equivalence of the new frame with the standard formalism}

\subsubsection{The Mathematical Basis of Statistical Mechanics}

Classical Statistical Mechanics is based on ordinary probability theory\cite
{probatheory} that employs a sample space $\mathcal{C}$ which is a
collection of sample points, regarded as mutually exclusive outcomes of a
hypothetical experiment. An event is a set of sample points, and the events,
under the operations of complementation, intersection and union form a
boolean algebra. The set $\mathcal{E}$ of events is called a $\sigma $
-field, and a probability law $\mu $ on $\mathcal{C}$ is a positive function
acting on events as prescribed by the mathematical theory of measure\cite
{measure} . Namely $\mu (\emptyset )=0$, $\mu (\mathcal{C})=1$, and for any
family $\left\{ A_{i}\right\} $ of disjoint events ($A_{i}\cap
A_{j}=\emptyset $ for $i\neq j$) we have $\mu (\cup _{i}A_{i})=\sum_{i}\mu
(A_{i})$. The probability $\mu (A)$ of an event ''$A$'' is the probability
that the outcome $x$ of some experiment belongs to $A$.

In our case, the sample space $\mathcal{C}$ is phase space and a sample
point is a state $(\overrightarrow{p},\overrightarrow{q})$. The set $%
\mathcal{E}$ of events is given by the family $\mathcal{B}(\mathcal{C})$ of
Borel sets in $\mathcal{C}$. Each probability law $\mu $ is associated with
a positive density $\rho (\overrightarrow{p},\overrightarrow{q})$ such that $%
\mu (A)=\int_{A}\rho d^{3}\overrightarrow{p}d^{3}\overrightarrow{q}$.

\subsubsection{Equivalence of Primary Axioms}

\begin{itemize}
\item  Now let us call $\Omega $ the set of orthogonal projectors of $%
\mathcal{H}=L^{2}(\mathcal{C})$ and define the mapping $\Pi $ from the set
of events $\mathcal{E}$ to $\Omega $ by: 
\begin{equation}
\forall A\in \mathcal{E}\text{, }\Pi (A)=\int d^{3}\overrightarrow{p}d^{3}%
\overrightarrow{q}\chi _{A}(\overrightarrow{p},\overrightarrow{q})|%
\overrightarrow{p},\overrightarrow{q}><\overrightarrow{p},\overrightarrow{q}|
\end{equation}
where $\chi _{A}(\overrightarrow{p},\overrightarrow{q})$ is the
characteristic function of the set $A$ ($\chi _{A}=1$ when $(\overrightarrow{%
p},\overrightarrow{q})$ belongs to $A$, and $\chi _{A}=0$ elsewhere).\newline
The family $\left\{ \Pi (A)\right\} _{A\in \mathcal{E}}$ gives a
representation of the boolean algebra of events in the quantum mathematical
frame as an abelian algebra of orthogonal projectors (and the $\Pi (A)$
become classical observables according to our previous definition).

\item  In ordinary formalism a particle is assumed to be represented by a
point $(\overrightarrow{p},\overrightarrow{q})$ at each time; but a true
physical experiment contains always some uncertainty, so the true
mathematical representation of a physical result is an expectation value
associated with some probability density $\rho $. Of course in principle, $%
\rho $ can be as close as you want to a ''$\delta $'', but ''$\delta $'' is
never reached. This means that we can never really know that a particle is
at the point $(\overrightarrow{p},\overrightarrow{q})$. Of course it is a
natural extrapolation, but not logically necessary. It is much more
realistic to say that a single particle is represented by a true (sharp)
density $\rho $. In the language of statistical sets, this means that we
look at a single particle as an element of the class of all particles
''prepared'' in the same way, the preparation being such that the
uncertainty on $\overrightarrow{p}$ and $\overrightarrow{q}$ can be reduced
(as much as you want), but cannot be cancelled.\newline
If we start with this new hypothesis, this means that now, a single particle
always exhibits some uncertainty. Moreover this hypothesis allows to
associate to each particle the object $\phi (\overrightarrow{p},%
\overrightarrow{q})=\sqrt{\rho (\overrightarrow{p},\overrightarrow{q})}$
that defines a normalized wave function of $\mathcal{H}$.\newline
This result can be also found using some extremum property. In ordinary
formalism, the one-point sets $\{(\overrightarrow{p},\overrightarrow{q})\}$
can be seen as the smallest (more precise) events for the ordering relation
''$\subset $'' on sets. Using the mapping $\Pi $ on events, we see that $%
A\subset B\Leftrightarrow \Pi (A)\leq \Pi (B)$ where ''$\leq $'' is the
inegality between self-adjoint operators. So a new representation of points
can be given by using $\Pi $ directly or using the extremum property. Since
the projector associated to a point $x$ is $\Pi (\{x\})=0$, a point $x$
cannot be associated with a true projector, and then we can only use the
extremum property to look for the minimal orthogonal projectors. Of course
these minimal projectors are given by $|\phi ><\phi |$, where $\phi $ is a
normalized vector in $\mathcal{H}$. So we recover that the best precise
events are associated with a normalized vector $\phi $ in $\mathcal{H}$, and
then a particle must be described by $\phi $.\newline
To finish, we remark that the real equivalent of the point $(\overrightarrow{%
p},\overrightarrow{q})$ is the unbounded state $|\overrightarrow{p},%
\overrightarrow{q}>$ or pseudo-projector $|\overrightarrow{p},%
\overrightarrow{q}><\overrightarrow{p},\overrightarrow{q}|$. But it cannot
be seen as a ''true state'' in this formalism.

\item  Now if we look at observables, each classical quantity $f(%
\overrightarrow{p},\overrightarrow{q})$ can be approximated by step
functions $\sum_{i}f_{i}\chi _{A_{i}}$ where $\chi _{A_{i}}$ are
characteristic functions of sets $A_{i}$. So we can extend by linearity the
mapping $\Pi $ to lift up any classical observable into its quantum
representation as 
\begin{equation}
\mathbf{f}=\int d^{3}\overrightarrow{p}d^{3}\overrightarrow{q}f(%
\overrightarrow{p},\overrightarrow{q})|\overrightarrow{p},\overrightarrow{q}%
><\overrightarrow{p},\overrightarrow{q}|
\end{equation}
Of course we recover the ''axiom of classicity'' and the fact that the
possible values of the observable are its eigen values.
\end{itemize}

\subsubsection{Equivalence of Statistical Axioms}

\begin{itemize}
\item  As seen before, a probability law on the ordinary frame is given by a
mapping $\mu $ acting on events following some precise rules. If we look for
a mapping $\mu _{Q}$ acting on the representation of events (that is on
orthogonal projectors) with the same rules, a theorem due to Gleason\cite
{gleason} proves that $\mu _{Q}$ is given by a density operator $\mathbf{D}$
such that $\mu _{Q}(\Pi (A))=Tr(\mathbf{D}.\Pi (A))$. This means that we can
associate to each classical probability law $\mu $ a density operator $%
\mathbf{D}$ such that $\mu (A)=Tr(\mathbf{D}.\Pi (A))$. If we develop this
formula we find: 
\begin{equation}
\mu (A)=Tr(\mathbf{D}.\Pi (A))=\int_{A}d^{3}\overrightarrow{p}d^{3}%
\overrightarrow{q}<\overrightarrow{p},\overrightarrow{q}|\mathbf{D}|%
\overrightarrow{p},\overrightarrow{q}>
\end{equation}
So the classical density $\rho (\overrightarrow{p},\overrightarrow{q})$ is
given by the diagonal element $\rho (\overrightarrow{p},\overrightarrow{q})=$
$<\overrightarrow{p},\overrightarrow{q}|\mathbf{D}|\overrightarrow{p},%
\overrightarrow{q}>$ (this quantity is positive since $\mathbf{D}$ is a
positive operator).\newline
Now for a particle ''in the state $\phi $'', if we assume that $\mathbf{D}%
=|\phi ><\phi |$, the corresponding classical density $\rho $ is given by: 
\begin{equation}
\rho (\overrightarrow{p},\overrightarrow{q})=<\overrightarrow{p},%
\overrightarrow{q}|\mathbf{D}|\overrightarrow{p},\overrightarrow{q}>=|<%
\overrightarrow{p},\overrightarrow{q}|\phi >|^{2}
\end{equation}
So we recover the formula introduced for justifying the existence of states.

\item  Now, if we use the relation $\rho (\overrightarrow{p},\overrightarrow{%
q})=<\overrightarrow{p},\overrightarrow{q}|\mathbf{D}|\overrightarrow{p},%
\overrightarrow{q}>$ for computing classical expectation value, we find: 
\begin{equation}
<f>=\int d^{3}\overrightarrow{p}d^{3}\overrightarrow{q}\rho f=Tr(\mathbf{D}.%
\mathbf{f})
\end{equation}
\emph{Remark:}\newline
We see that it is impossible to distinguish the statistical effects due to a
general density $\mathbf{D}$ from those due to a pure state, because only
the diagonal elements $<\overrightarrow{p},\overrightarrow{q}|\mathbf{D}|%
\overrightarrow{p},\overrightarrow{q}>$ are relevant in our case: the pure
state $<\overrightarrow{p},\overrightarrow{q}|\phi >=\sqrt{<\overrightarrow{p%
},\overrightarrow{q}|\mathbf{D}|\overrightarrow{p},\overrightarrow{q}>}$
gives exactly the same results than $\mathbf{D}$ (in our classical frame).
So pure states and general densities $\mathbf{D}$ are different on a logical
level, but they are physically undistinguishable.
\end{itemize}

\subsubsection{Equivalence of Collapse Axiom with Conditional Probability}

In ordinary probability theory, we use conditional probability to take into
account new results to modify a probability law. Namely, if we have a
probability law $\mu $ and if we know that some event $E$ is realized, the
new probability law $\mu _{E}$ is: 
\begin{equation}
\mu _{E}(A)=\frac{\mu (A\cap E)}{\mu (E)}
\end{equation}

In our case, if $\mu $ is associated with a density $\mathbf{D}$ we have: 
\begin{equation}
\mu _{E}(A)=\frac{Tr(\mathbf{D}.\Pi (A\cap E))}{Tr(\mathbf{D}.\Pi (E))}
\end{equation}

If we use the fact that $\Pi (A\cap E)=\Pi (A).\Pi (E)=\Pi (E).\Pi (A)$, we
find: 
\begin{equation}
\mu _{E}(A)=\frac{Tr(\Pi (E).\mathbf{D}.\Pi (E)\Pi (A))}{Tr(\mathbf{D}.\Pi
(E))}
\end{equation}

If we introduce the operator $\mathbf{D}_{E}$ defined as: 
\begin{equation}
\mathbf{D}_{E}=\frac{1}{Tr(\mathbf{D}.\Pi (E))}\Pi (E).\mathbf{D}.\Pi (E)
\end{equation}
$\mathbf{D}_{E}$ is a true density operator and $\mu _{E}(A)=Tr(\mathbf{D}
_{E}.\Pi (A))$. So $\mathbf{D}_{E}$ is the density operator associated with $%
\mu _{E}$.\newline
So, we find that conditional probability is implemented in the quantum frame
with the formula of the ''collapse axiom''. Now, let apply this result to
the case of some measure of a classical observable $F(\overrightarrow{p},%
\overrightarrow{q})$ that specifies that the possible numerical outcomes $f$
belong to a set $V\subset \mathbf{R}$. This means that after the measure, we
know that the particle ''is'' in the region $F^{-1}(V)$ of phase space. This
means that we know that the event $F^{-1}(V)$ is realized. So the initial
probability law $\mu $ associated with the density $\mathbf{D}$ must be
changed and the new density $\mathbf{D}^{\prime }$ is: 
\begin{equation}
\mathbf{D}^{\prime }=\frac{1}{Tr(\mathbf{D}.\mathbf{P}(V))}\mathbf{P}(V).%
\mathbf{D}.\mathbf{P}(V)\text{ with }\mathbf{P}=\Pi \circ F^{-1}
\end{equation}
It is not hard to see that $\mathbf{P}=\Pi \circ F^{-1}$ is exactly the
Projection Valued Measure associated with the quantum version $\mathbf{F}$
of the classical observable $F$. This shows that the collapse axiom is (in
our case) a translation of conditional probability.

\subsubsection{Axiom Of Evolution}

We will see in the following that this axiom allows to recover Classical
Dynamics.

\subsection{Conclusion}

This new mathematical framework for classical Mechanics is completely
equivalent to the ordinary one to do Physics. But, while the usual frame is
mathematically and physically closed (all things that you can write possess
a classical meaning), now this new formalism can be extended by removing
some axiom, more specially the ''axiom of classicity'', by assuming that,
perhaps, some unclassical observables also have a physical meaning. How we
will see later, Quantization corresponds to such a process. We finish with
some remarks.

\begin{itemize}
\item  Interference Effect\newline
Usually, interferences are seen as one of the main consequences of quantum
formalism. We want to point out that they cannot be observed in our case,
thanks to our ''axiom of classicity''. Actually to see interferences, we
need at least two physical observables that are not diagonal in the same
basis, and this case is excluded by our axiom. This can be seen also by the
fact that pure states cannot be distinguished from general densities.

\item  Delocalization of States\newline
We have introduced the state $\phi $ that describes a particle by assuming
that $\rho (\overrightarrow{p},\overrightarrow{q})=|<\overrightarrow{p},%
\overrightarrow{q}|\phi >|^{2}$ is a sharp density in phase space. But
nothing forbids $\phi $ to be very delocalized, and in principle, we must
look at this delocalization as being intrinsic (not due to a lack of
knowledge as in the case of a general density $\mathbf{D}$). So the question
that arises is: why classical objects must be described in practice with
only very sharp densities? The answer to this question is contained in our
previous remark on the physical impossibility (in this frame) to distinguish
states from general densities (the physical effects of a density can be
always represented by a state). So, it is only a matter of convenience (and
consistency) to decide that ''true states'' correspond to very small
delocalization. Of course, it must be also possible to develop some
argumentation based on ''decoherence''.

\item  Local $U(1)$ invariance\newline
Since the computation of any physical quantity only depends on $\rho (%
\overrightarrow{p},\overrightarrow{q})=|<\overrightarrow{p},\overrightarrow{q%
}|\phi >|^{2}$, we can modify the state $\phi (\overrightarrow{p},%
\overrightarrow{q})$ by any phase factor $\exp (i\theta (\overrightarrow{p},%
\overrightarrow{q}))$ without changing physical results.\bigskip 
\end{itemize}

Now, we can analyze how the classical structure of Poisson Brackets can be
implemented in this new frame, and how we can recover Classical Dynamics.

\section{Representation of Poisson Brackets in the New Framework}

Let $f(\overrightarrow{p},\overrightarrow{q})$ an observable (real
function). We associate to $f$ a self adjoint operator $\mathbf{X}_{f}$
acting on states $\phi \in \mathcal{H}$ by: 
\begin{equation}
\mathbf{X}_{f}\phi =-i\left\{ f,\phi \right\}
\end{equation}

$\mathbf{X}_{f}$ defines a generator of a one-parameter unitary group $%
\mathbf{U}_{\alpha }$ with: 
\begin{equation}
\mathbf{U}_{\alpha }=\exp [-i\alpha \mathbf{X}_{f}]
\end{equation}

And if we define the state $|\phi _{\alpha }>$ $=\mathbf{U}_{\alpha }|\phi
>, $ we have: 
\begin{equation}
i\frac{\partial }{\partial \alpha }|\phi _{\alpha }>\text{ }=\mathbf{X}
_{f}|\phi _{\alpha }>\text{ or }\frac{\partial }{\partial \alpha }\phi
_{\alpha }=-\left\{ f,\phi _{\alpha }\right\}
\end{equation}

Now it is not hard to see that $\mathbf{U}_{\alpha }$ acts on unbounded
states $|\overrightarrow{p},\overrightarrow{q}>$ exactly as the classical
transformation (\ref{Generalpqevol}): 
\begin{equation}
\mathbf{U}_{\alpha }|\overrightarrow{p},\overrightarrow{q}>\text{ }=|%
\overrightarrow{p_{\alpha }},\overrightarrow{q_{\alpha }}>  \label{Uamap}
\end{equation}
where $(\overrightarrow{p_{\alpha }},\overrightarrow{q_{\alpha }})$ is the
trajectory with initial conditions $(\overrightarrow{p},\overrightarrow{q})$.

So $\mathbf{U}_{\alpha }$ maps states into states.

Now, if we consider another classical observable $g$, and the associated
operator $\mathbf{g}$, we can define the transformed operator $\mathbf{g}
_{\alpha }=\mathbf{U}_{\alpha }^{\dagger }\mathbf{gU}_{\alpha }$.

Because (\ref{Uamap}), $\mathbf{g}_{\alpha }$ is also a self adjoint
operator associated with a classical observable $g_{\alpha }$ and we have: 
\begin{equation}
\frac{\partial }{\partial \alpha }\mathbf{g}_{\alpha }=i[\mathbf{X}_{f},%
\mathbf{g}_{\alpha }]\text{ or }\frac{\partial }{\partial \alpha }g_{\alpha
}=\left\{ f,g_{\alpha }\right\}
\end{equation}

To end, the commutator $[\mathbf{X}_{f},\mathbf{X}_{g}]$ of two operators $%
\mathbf{X}_{f}$ and $\mathbf{X}_{g}$ is given by: 
\begin{equation}
i[\mathbf{X}_{f},\mathbf{X}_{g}]=\mathbf{X}_{\{f,g\}}  \label{PBcomm}
\end{equation}

So the linear mapping $f\rightarrow \mathbf{X}_{f}$ is a representation of
the action induced by $f$ through Poisson Brackets. Moreover (\ref{PBcomm})
shows that we have also a natural representation of $\left\{ .,.\right\} $
as $i\left[ .,.\right] $.

Nevertheless, since $\mathbf{X}_{f}=0$ for $f=constant$, we cannot recover
the special value $\left\{ p_{i},q_{j}\right\} =\delta _{ij}$.

\textbf{Remark on the ''Active'' and ''Passive'' representations of
Observables}\newline
From general quantum axiomatics, each $\mathbf{X}_{f}$ as a self-adjoint
operator defines mathematically a possible observable (even if physically $%
\mathbf{X}_{f}$ is not a classical object because our axiom of classicity).
Then, on a purely mathematical point of view, we have two possible
self-adjoint operators deducible from the classical quantity $f(x)$:

\begin{itemize}
\item  $\mathbf{f}=\int dx$ $f(x)|x><x|$

\item  $\mathbf{X}_{f}$.
\end{itemize}

$\mathbf{f}$ defines the observable as a ''passive object'', that is a data;
while $\mathbf{X}_{f}$ is the ''active version'' of $f$ as the action
induced by $f$ through Poisson Brackets. Moreover $\mathbf{f}$ and $\mathbf{X%
}_{f}$ are independent, because $[\mathbf{f},\mathbf{X}_{f}]=0$.

Then, while in the ''passive representation'' all classical observables $%
\mathbf{f}$ commute, in the ''active representation'' commutators are
related to Poisson brackets (equation \ref{PBcomm}).

So, from the mathematical point of view, $\mathbf{X}_{f}$ is an alternative
representation of the classical quantity $f(x)$. Unfortunately on a physical
level,\textbf{\ }$\mathbf{f}$ and $\mathbf{X}_{f}$ have not the same
homogeneity: $\mathbf{X}_{f}$ is homogeneous to $\mathbf{f}$ divided by an
action. So, as long as we have not a specific unit of action, we cannot
follow up this process.

To end this remark, we want to indicate that this idea of representing
classical observables by a pair of operators has already been used in the
''superoperator formalism'' by C. George and I. Prigogine\cite{prigogine2} .

\section{Classical Dynamics in the New Framework}

\subsection{New Formulation of Equations of Motion}

From the general axioms, we know that the evolution of states is given by
the unitary operators $\mathbf{U}_{t,t_{0}}$ such that $\phi _{t}=\mathbf{U}
_{t,t_{0}}(\phi _{t_{0}})$. Moreover the previous paragraphs II-B1 specify
that Dynamics is defined by the data of the classical Hamiltonian $H$
(eventually time-dependent), and we have seen in II-B2,3 that ''weak
dynamical equations'' (\ref{meanhamilequs}) are more general formula to
specify dynamics. So, we assume in the following that (\ref{meanhamilequs})
are our basic equations of motion.\newline
Since we have seen in III-B3 the equivalence between quantum and classical
formula for expectation values of classical observables, using the classical
density $\rho (\overrightarrow{p},\overrightarrow{q},t)=|<\overrightarrow{p},%
\overrightarrow{q}|\phi _{t}>|^{2}$; equations (\ref{meanhamilequs}) can be
written: 
\begin{equation}
\left\{ 
\begin{array}{l}
\frac{d}{dt}<\phi _{t}|\overrightarrow{\mathbf{q}}|\phi _{t}>=<\phi _{t}|%
\overrightarrow{\nabla _{p}}H(\overrightarrow{\mathbf{p}},\overrightarrow{%
\mathbf{q}},t)|\phi _{t}> \\ 
\frac{d}{dt}<\phi _{t}|\overrightarrow{\mathbf{p}}|\phi _{t}>=-<\phi _{t}|%
\overrightarrow{\nabla _{q}}H(\overrightarrow{\mathbf{p}},\overrightarrow{%
\mathbf{q}},t)|\phi _{t}>
\end{array}
\right.  \label{newweakdynamics}
\end{equation}
where $\overrightarrow{\nabla _{p}}H$ and $\overrightarrow{\nabla _{q}}H$
can be seen as functions of the operators $\overrightarrow{\mathbf{p}}$ and $%
\overrightarrow{\mathbf{q}}$, since $\overrightarrow{\mathbf{p}}$ and $%
\overrightarrow{\mathbf{q}}$ commute.

Equations (\ref{newweakdynamics}) are our new basic dynamical equations.

\subsection{Classical Dynamics}

\subsubsection{Primary Equations}

Any operator $\mathbf{f}$ associated with the classical observable $f(%
\overrightarrow{p},\overrightarrow{q})$ can be written as $\mathbf{f=}f(%
\overrightarrow{\mathbf{p}},\overrightarrow{\mathbf{q}})$, since $%
\overrightarrow{\mathbf{p}}$ and $\overrightarrow{\mathbf{q}}$ commute. So
the expectation value $<\mathbf{f}>_{t}$ verifies: 
\begin{equation}
<\mathbf{f}>_{t}=<\phi _{t}|f(\overrightarrow{\mathbf{p}},\overrightarrow{%
\mathbf{q}})|\phi _{t}>=<\phi _{0}|f(U_{t,t_{0}}^{\dagger }\overrightarrow{%
\mathbf{p}}U_{t,t_{0}},U_{t,t_{0}}^{\dagger }\overrightarrow{\mathbf{q}}
U_{t,t_{0}})|\phi _{0}>
\end{equation}

Now we introduce the operators $\overrightarrow{\mathbf{q}}(t)$ and $%
\overrightarrow{\mathbf{p}}(t)$ defined as: 
\begin{equation}
\left\{ 
\begin{array}{l}
\overrightarrow{\mathbf{q}}(t)=U_{t,t_{0}}^{\dagger }\overrightarrow{\mathbf{%
\ \ q}}U_{t,t_{0}} \\ 
\overrightarrow{\mathbf{p}}(t)=U_{t,t_{0}}^{\dagger }\overrightarrow{\mathbf{%
\ \ p}}U_{t,t_{0}}
\end{array}
\right.  \label{operators(t)}
\end{equation}

Equations for expectation values (\ref{newweakdynamics}) become: 
\begin{equation}
\left\{ 
\begin{array}{l}
<\phi _{0}|\frac{d}{dt}\overrightarrow{\mathbf{q}}(t)|\phi _{0}>\text{ }=%
\text{ }<\phi _{0}|\overrightarrow{\nabla _{p}}H(\overrightarrow{\mathbf{p}}
(t),\overrightarrow{\mathbf{q}}(t),t)|\phi _{0}> \\ 
<\phi _{0}|\frac{d}{dt}\overrightarrow{\mathbf{p}}(t)|\phi _{0}>\text{ }
=-<\phi _{0}|\overrightarrow{\nabla _{q}}H(\overrightarrow{\mathbf{p}}(t),%
\overrightarrow{\mathbf{q}}(t),t)|\phi _{0}>
\end{array}
\right.  \label{meanequs1}
\end{equation}

Since equations (\ref{meanequs1}) \textbf{must be valid for any state} $\phi
_{0}$, we deduce the following equations of evolution for the operators $%
\overrightarrow{\mathbf{q}}(t)$ and $\overrightarrow{\mathbf{p}}(t)$: 
\begin{equation}
\left\{ 
\begin{array}{l}
\frac{d}{dt}\overrightarrow{\mathbf{q}}(t)=\overrightarrow{\nabla _{p}}H(%
\overrightarrow{\mathbf{p}}(t),\overrightarrow{\mathbf{q}}(t),t) \\ 
\frac{d}{dt}\overrightarrow{\mathbf{p}}(t)=-\overrightarrow{\nabla _{q}}H(%
\overrightarrow{\mathbf{p}}(t),\overrightarrow{\mathbf{q}}(t),t)
\end{array}
\right.  \label{equsforevolutionObservables}
\end{equation}

(\ref{equsforevolutionObservables}) are the natural formulation of equations
of motion in the new framework, and they can be symbolically identified with
the Hamiltonian equations (\ref{hamilequs}).

\subsubsection{Operator of Evolution and Liouville Equation}

Now if we look at the operator $\mathbf{X}_{H}(t)$ as defined in III-C: 
\begin{equation}
\mathbf{X}_{H}(t)=-i\left( \overrightarrow{\nabla _{p}}H\overrightarrow{
.\nabla _{q}}-\overrightarrow{\nabla _{q}}H.\overrightarrow{\nabla _{p}}
\right)
\end{equation}

We have: 
\begin{equation}
\left\{ 
\begin{array}{l}
\lbrack \mathbf{X}_{H}(t),\overrightarrow{\mathbf{q}}]=-i\overrightarrow{
\nabla _{p}}H(\overrightarrow{\mathbf{p}},\overrightarrow{\mathbf{q}},t) \\ 
\lbrack \mathbf{X}_{H}(t),\overrightarrow{\mathbf{p}}]=i\overrightarrow{
\nabla _{q}}H(\overrightarrow{\mathbf{p}},\overrightarrow{\mathbf{q}},t)
\end{array}
\right.
\end{equation}

Let us define $\mathbf{U}_{t,t_{0}}$ as the unitary operator generated by $%
\mathbf{X}_{H}(t)$: 
\begin{equation}
\left\{ 
\begin{array}{l}
i\frac{d}{dt}\mathbf{U}_{t,t_{0}}=\mathbf{X}_{H}(t)\mathbf{U}_{t,t_{0}} \\ 
\mathbf{U}_{t_{0},t_{0}}=\mathbf{1}
\end{array}
\right.
\end{equation}

A simple checking shows that now the operators $\overrightarrow{\mathbf{q}}
(t)$ and $\overrightarrow{\mathbf{p}}(t)$ defined by (\ref{operators(t)})
verify equations (\ref{equsforevolutionObservables}) and then $\mathbf{U}
_{t,t_{0}}$ is the operator of evolution. We deduce that for any state $\phi
_{0}$ at time $t_{0}$ we have: 
\begin{equation}
\phi (t)=\mathbf{U}_{t,t_{0}}(\phi _{0})
\end{equation}

Then using the definition of $\mathbf{U}_{t,t_{0}}$, the wave function $\phi
(\overrightarrow{p},\overrightarrow{q},t)$ verifies Liouville equation: 
\begin{equation}
\frac{\partial \phi }{\partial t}=-\{H,\phi \}
\end{equation}

Because this differential equation is of order one with real coefficients,
we recover that the probability density $\rho (\overrightarrow{p},%
\overrightarrow{q},t)=\left| \phi (\overrightarrow{p},\overrightarrow{q}
,t)\right| ^{2}$ must also verify Liouville equation. Now, since any density
operator $\mathbf{D}$ has the splitting $\mathbf{D}=\sum_{n}p_{n}|\phi
_{n}><\phi _{n}|$, we recover that any classical density $\rho (%
\overrightarrow{p},\overrightarrow{q},t)=$ $<\overrightarrow{p},%
\overrightarrow{q}|D_{t}|\overrightarrow{p},\overrightarrow{q}>$ must verify
also Liouville equation.

We see that only starting from equations for expectation values, and even if
''$\delta $'' densities are excluded, we recover completely the equations of
standard Classical Mechanics. This result is due to the fact that we can
find densities as close as we want to ''$\delta $''.

\section{The Quantization Process}

\subsection{The problem of Unclassical Observables: what is Quantization?}

Our frame of Classical Mechanics is based on the ''axiom of classicity''
that limits the range of physical observables. We analyze here the
possibility of giving a physical meaning to some unclassical observable (and
then breaking down our axiom).

The starting point of our procedure is the representation of classical
events given by our mapping $\Pi $. As we can verify easily, for each
classical event ''$A$'' that gives a non-zero projector $\Pi (A)$, $\Pi (A)$
corresponds in fact to an infinite dimensional subspace of $\mathcal{H}$.
This means that the one-dimensional projectors $\mathbf{\pi }_{\phi }=|\phi
><\phi |$ that represents the ''minimal events'' and define states, are not
classical events. Since we only possess the projectors $\Pi (A)$ to select
particles, we can never decide if a particle is in a given state $\phi $
(this is because we cannot distinguish states from general densities). So in
this classical picture, a particle is described by a mathematical object
that cannot be completely specified from experiment (exactly as points in
phase space that cannot be experimently reached). But this means that we
cannot specify any more what is really the full space $\mathcal{S}$ of all $%
\phi $. So we can imagine two different situations:\newline
- The space $\mathcal{S}=\mathcal{H}$, and then all states are possible%
\newline
- The space $\mathcal{S}$ is only a proper subspace of $\mathcal{H}$.\newline
In this last case, we can define the orthogonal projector $\mathbf{\pi }_{%
\mathcal{S}}$ on $\mathcal{S}$. By definition $\mathbf{\pi }_{\mathcal{S}}$
cannot be a classical event, but nevertheless $\mathbf{\pi }_{\mathcal{S}}$
has a physical meaning as the projector on all physical states. So we can
give a physical meaning to an unclassical object and moreover this
unclassical event must be always realized since all physical states $\phi $
must verify $\mathbf{\pi }_{\mathcal{S}}\phi =\phi $.\newline
Another way to present the problem is to look at the full set of orthogonal
projectors as the set of all symbolic logical questions on a system. In the
frame of Classical Mechanics, only a subset of questions (the projectors $%
\Pi (A)$) possesses a physical answer (true or false). The remainder must be
either unphysical questions, either \emph{undecidable} questions. The
''question $\mathbf{\pi }_{\mathcal{S}}$'' is undecidable.\newline
So we have the opportunity of ''creating'' a new logical frame by
postulating that some ''question $\mathbf{\pi }_{\mathcal{S}}$'' has a
positive answer. Of course, we need physical arguments to choose $\mathbf{\
\pi }_{\mathcal{S}}$ in order to preserve basic physical results.\newline
In our point of view, Quantization corresponds to this kind of
postulate.\bigskip

In the following, we assume that some proper subspace $\mathcal{S}$ of $%
\mathcal{H}$ (represented by $\mathbf{\pi }_{\mathcal{S}}$) contains all
possible states.\newline
We analyze first the consequences of this postulate and then we give the
arguments to choose $\mathbf{\pi }_{\mathcal{S}}$.

\subsection{Consequences of a Quantization}

Of course, we must assume that all general axioms listed in III-A1,2,3,4 are
always valid, but on the Hilbert space $\mathcal{S}$. Moreover, we must
analyze apart how to take into account our axiom of classicity.

First of all, since the basic objects of our formalism are now operators on $%
\mathcal{S}$ which is a subspace of $\mathcal{H}$, we want to specify the
relation between self-adjoint operators on $\mathcal{H}$ and self-adjoint
operators on $\mathcal{S}$. Since $\mathbf{\pi }_{\mathcal{S}}$ is a
projector, a self-adjoint operator $\mathbf{a}$ on $\mathcal{H}$ corresponds
to a self-adjoint operator on $\mathcal{S}$ if and only if: 
\begin{equation}
\mathbf{\pi }_{\mathcal{S}}.\mathbf{a}.\mathbf{\pi }_{\mathcal{S}}=\mathbf{a}
\label{projoperator}
\end{equation}
So this relation (\ref{projoperator}) must now be verified for both
observables and densities.\newline
Now we analyze how classical observables are modified.

\subsubsection{The ''Correspondence Principle''}

Let $f$ a classical observable and $\mathbf{f}=\int f(x)|x><x|dx$ the
corresponding operator. For any density $\mathbf{D}$, we have seen that the
expectation value of $f$ is given by $<f>=Tr(\mathbf{D}.\mathbf{f})$. But
now the only possible densities verify equation (\ref{projoperator}) and
then $<f>$ can be written as $<f>=Tr(\mathbf{D}.\mathbf{\pi }_{\mathcal{S}}.%
\mathbf{f}.\mathbf{\pi }_{\mathcal{S}})$. So in any physical situation, we
only need the self-adjoint operator $\mathbf{f}_{\mathcal{S}}=\pi _{\mathcal{%
\ S}}\mathbf{f}\pi _{\mathcal{S}}$, and by construction $\mathbf{f}_{%
\mathcal{S}}$ defines a true observable on the physical Hilbert space $%
\mathcal{S}$.\newline
\emph{So it is natural to assume that the operator} $\mathbf{f}_{\mathcal{S}%
}=\mathbf{\pi }_{\mathcal{S}}.\mathbf{f.\pi }_{\mathcal{S}}$ \emph{defines
the new version of the classical observable} $\mathbf{f}$: \emph{we call} $%
\mathbf{f}_{\mathcal{S}}$ \emph{the quantized observable associated with }$%
\mathbf{f}$.\newline
This allows to justify and not to postulate the famous ''correspondence
principle'' that gives quantized version of classical observable. We will
see later that \emph{almost} all these quantized observables correspond
effectively to usual quantum operators.

\subsubsection{Remarks on quantized Observables}

Of course, the simplest quantized observables are $\mathbf{\pi }_{\mathcal{S}%
}\overrightarrow{\mathbf{p}}\mathbf{\pi }_{\mathcal{S}}$ and $\mathbf{\pi }_{%
\mathcal{S}}\overrightarrow{\mathbf{q}}\mathbf{\pi }_{\mathcal{S}}$. Since $%
\overrightarrow{\mathbf{p}}$ and $\overrightarrow{\mathbf{q}}$ are the basic
objects of our classical frame, $\mathbf{\pi }_{\mathcal{S}}\overrightarrow{%
\mathbf{p}}\mathbf{\pi }_{\mathcal{S}}$ and $\mathbf{\pi }_{\mathcal{S}}%
\overrightarrow{\mathbf{q}}\mathbf{\pi }_{\mathcal{S}}$ must become the new
fundamental observables; and these new operators do not commute in general.%
\newline
In the case of more complex classical observables, it can exist some
ambiguity in the definition of the quantized version.\newline
For example, if we look at the quantized observable $\overrightarrow{\mathbf{%
L}}$ associated with the classical orbital momentum $\overrightarrow{l}=%
\overrightarrow{p}\wedge \overrightarrow{q}$, we can say that $%
\overrightarrow{\mathbf{L}}=\overrightarrow{\mathbf{L}_{1}}=\mathbf{\pi }_{%
\mathcal{S}}\overrightarrow{\mathbf{p}}\wedge \overrightarrow{\mathbf{q}}%
\mathbf{\pi }_{\mathcal{S}}$ according to our procedure, but we can also
define $\overrightarrow{\mathbf{L}}=\overrightarrow{\mathbf{L}_{2}}=\frac{1}{%
2}(\mathbf{\pi }_{\mathcal{S}}\overrightarrow{\mathbf{p}}\mathbf{\pi }_{%
\mathcal{S}}\wedge \mathbf{\pi }_{\mathcal{S}}\overrightarrow{\mathbf{q}}%
\mathbf{\pi }_{\mathcal{S}}-\mathbf{\pi }_{\mathcal{S}}\overrightarrow{%
\mathbf{q}}\mathbf{\pi }_{\mathcal{S}}\wedge \mathbf{\pi }_{\mathcal{S}}%
\overrightarrow{\mathbf{p}}\mathbf{\pi }_{\mathcal{S}})$. In both cases we
have a well-defined self-adjoint operator, and then a possible observable
associated with the classical quantity $\overrightarrow{l}$ (in general $%
\overrightarrow{\mathbf{L}_{1}}$ and $\overrightarrow{\mathbf{L}_{2}}$
define two different operators). Of course only $\overrightarrow{\mathbf{L}%
_{1}}$ is a direct quantization (according to our procedure); but it is not
sufficient to prove that $\overrightarrow{\mathbf{L}_{1}}$ is the right
answer, because orbital momentum is classically the generator of rotation
and we can demand that this general physical property remains unchanged.
This means that, for complex observables, we need generally some external
arguments on the physical role played by the observable (for example as
generator of symmetry), to decide which is the right quantized object. We
will see in section X-B, precisely in the case of $\overrightarrow{l}=%
\overrightarrow{p}\wedge \overrightarrow{q}$, that we must examine carefully
different possible observables to deduce the right one.

\subsubsection{Quantization of Classical Events}

As seen before, classical events are represented by the projectors $\Pi (A)$
. Our quantization transform $\Pi (A)$ into $\Pi _{\mathcal{S}}(A)=\mathbf{\
\pi }_{\mathcal{S}}\Pi (A)\mathbf{\pi }_{\mathcal{S}}$. But now $\Pi _{%
\mathcal{S}}(A)$ is not any more an orthogonal projector and cannot be
associated with a true ''logical observable''. $\Pi _{\mathcal{S}}(A)$ must
be interpreted as a ''semi-classical event'' or as a quasiprojector. This
label becomes more obvious if we remark that: 
\begin{equation}
\Pi _{\mathcal{S}}(A)=\int_{A}d^{3}\overrightarrow{p}d^{3}\overrightarrow{q}%
\mathbf{\pi }_{\mathcal{S}}|\overrightarrow{p},\overrightarrow{q}><%
\overrightarrow{p},\overrightarrow{q}|\mathbf{\pi }_{\mathcal{S}}
\end{equation}
while: 
\begin{equation}
\mathbf{1}_{\mathcal{S}}=\mathbf{\pi }_{\mathcal{S}}=\int d^{3}%
\overrightarrow{p}d^{3}\overrightarrow{q}\mathbf{\pi }_{\mathcal{S}}|%
\overrightarrow{p},\overrightarrow{q}><\overrightarrow{p},\overrightarrow{q}|%
\mathbf{\pi }_{\mathcal{S}}
\end{equation}

The states $\left\{ |\xi _{\overrightarrow{p},\overrightarrow{q}}>=\mathbf{\
\pi }_{\mathcal{S}}|\overrightarrow{p},\overrightarrow{q}>\right\} $ are not
any more a ''continuous orthogonal basis'', but they verify the previous
closure relation on the physical Hilbert space $\mathcal{S}$. So these
states define in fact an overcomplete orthogonal basis and can be seen as
''semi-classical states''. We detail this point in section X-C,D.

Now, we can analyze what kind of arguments can be used to specify the
projector $\mathbf{\pi }_{\mathcal{S}}$.

\subsection{Our Arguments to choose a Quantization}

Our arguments for quantization are based on the physical existence of a
fundamental group of transformations acting on phase space, namely the
symmetry group. As indicated in paragraph II-C, the symmetry group is
Galileo group $G$ (up to some discrete transformations). So we look for a
representation of the group $G$ on the Hilbert space $\mathcal{H}$ in such a
way that classical objects are always transformed in the same way.

\subsubsection{How to represent Galileo group}

Classically an observable $f(\overrightarrow{p},\overrightarrow{q})$ is
transformed into $f_{u}=f\circ u^{-1}$ in a symmetry $u\in G$. Now, the
operator associated with $f_{u}$ is: 
\begin{equation}
\mathbf{f}_{u}=\int d^{3}\overrightarrow{p}d^{3}\overrightarrow{q}(f\circ
u^{-1})(\overrightarrow{p},\overrightarrow{q})|\overrightarrow{p},%
\overrightarrow{q}><\overrightarrow{p},\overrightarrow{q}|
\end{equation}
Since any symmetry $u$ preserves the element of volume in phase space, we
have: 
\begin{equation}
\mathbf{f}_{u}=\int d^{3}\overrightarrow{p}d^{3}\overrightarrow{q}f(%
\overrightarrow{p},\overrightarrow{q})|u(\overrightarrow{p},\overrightarrow{q%
})><u(\overrightarrow{p},\overrightarrow{q})|
\end{equation}

Then we can write: 
\begin{equation}
\mathbf{f}_{u}=\mathbf{T}_{u}.\mathbf{f}.\mathbf{T}_{u}^{\dagger }
\end{equation}
if we define the action of the operator $\mathbf{T}_{u}$ on $\mathcal{H}$
as: 
\begin{equation}
\mathbf{T}_{u}|\overrightarrow{p},\overrightarrow{q}>=\exp (i\theta _{u}(%
\overrightarrow{p},\overrightarrow{q}))|u(\overrightarrow{p},\overrightarrow{
q})>
\end{equation}
where $\theta _{u}(\overrightarrow{p},\overrightarrow{q})$ is an unspecified
phase factor (in fact we use the local $U(1)$ invariance).

This means that any projective representation of $G$ preserves classical
properties. Of course the choice $\theta _{u}=0$ gives a right
representation, but this solution is not necessarily the unique one.
Moreover, if there exists different non-equivalent representations, we have
to do a choice: if this picture is really the right mathematical framework,
we must say that only one of these representations is real.\newline
In the following, we assume that this choice has been done, and then the
functions $\theta _{u}$ are specified.

Now, we are ready to develop our arguments for quantization.

\subsubsection{Arguments for Quantization}

We remark first that ''$\emph{Phase}$ \emph{space is connected by symmetries}
'': given two points $x_{0}$ and $x_{1}$ in phase space, we can always find
some symmetry $u\in G$ such that $x_{1}=u(x_{0})$. So starting from any
point $x_{0}$, the action of the full symmetry group $G$ generates phase
space. This means that the set of physical states is generated by the
symmetry group: to specify dynamical properties of a particle, we don't need
a larger space than a space generated by $G$.\newline
Now in the new framework, the points $x$ of phase space are used to build
the ''continuous orthogonal basis'' $\left\{ |x>\right\} $; so this complete
orthogonal basis of $\mathcal{H}$ is generated by the action $G$ (or its
representation) on some ket $|x_{0}>$. But in general we can build others
(orthogonal) basis with this kind of property: we can find orthogonal basis $%
\left\{ |y,n>\right\} $ (where $y$ specifies continuous variables and $n$ is
an integer parameter), such that the action of the representation $\{\mathbf{%
T}_{u}\}$ on some ket $|y_{0},n>$ generates the full set $\left\{
|y,n>\right\} $ for a fixed $n$. So each subspace $\mathcal{H}_{n}$
associated with the projector $\Pi _{n}=\int dy|y,n><y,n|$ possesses exactly
the same property than the initial Hilbert space $\mathcal{H}$. Moreover two
states $\phi _{0}$ and $\phi _{1}$ belonging to two different subspaces $%
\mathcal{H}_{n_{0}}$ and $\mathcal{H}_{n_{1}}$ can never be connected by a
symmetry $\mathbf{T}_{u}$. So it is logical to assume that only one of these
subspaces $\mathcal{H}_{n}$ is sufficient to specify dynamical properties of
a particle.

\subsubsection{Conclusion}

If we demand that the Hilbert space of states verify the same property than
classical phase space, it is sufficient to assume that the physical Hilbert
space $\mathcal{S}$ is a subspace of irreducible representation of the $\{%
\mathbf{T}_{u}\}$: this specifies our quantization.\newline
So, logically we must first analyze all possible projective representations
of Galileo group, before looking for quantization. Nevertheless, since in
this article we want to focuss to the ''right choice'', we only develop in
the following paragraphs the intuitive arguments leading to the ''right''
representation compatible with Quantum Mechanics (there is no mathematical
reason to choose this special representation).

\section{Representations of Galileo Group}

As indicated at the end of the last paragraph, we have no mathematical
reason to choose a special representation; but the lack of mathematical
inference does not mean that we have no physical arguments. In fact, we will
prove in the following that assuming the physical existence of some ''unit
of action'' implies the existence of some new symmetry with no classical
equivalent. Taking into account this new symmetry, we can guess the
representation we look for.

To begin, we look first at the classical situation corresponding to a
cancellation of all the coefficients $\theta _{u}(x)$ introduced in section
VI.

\subsection{The classical representation}

We can use the generators $\mathbf{X}_{f}$ induced by Poisson Brackets (IV)
to build this classical representation of Galileo group.

The continuous transformations of the group are space translations, Galileo
boosts and rotations that are respectively associated with the observables $%
\overrightarrow{p}$, $\overrightarrow{q}$ and $\overrightarrow{l}=%
\overrightarrow{q}\wedge \overrightarrow{p}$. Namely, using the notations of
section IV: 
\begin{equation}
\left\{ 
\begin{array}{l}
\exp \left[ -i\overrightarrow{q_{0}}.\mathbf{X}_{\overrightarrow{p}}\right] |%
\overrightarrow{p},\overrightarrow{q}>\text{ }=|\overrightarrow{p},%
\overrightarrow{q}+\overrightarrow{q_{0}}> \\ 
\exp \left[ i\overrightarrow{p_{0}}.\mathbf{X}_{\overrightarrow{q}}\right] |%
\overrightarrow{p},\overrightarrow{q}>\text{ }=|\overrightarrow{p}+%
\overrightarrow{p_{0}},\overrightarrow{q}> \\ 
\exp \left[ -i\overrightarrow{\omega }.\mathbf{X}_{\overrightarrow{l}
}\right] |\overrightarrow{p},\overrightarrow{q}>\text{ }=|\mathcal{R}_{%
\overrightarrow{\omega }}(\overrightarrow{p}),\mathcal{R}_{\overrightarrow{
\omega }}(\overrightarrow{q})>
\end{array}
\right.
\end{equation}
where $\mathcal{R}_{\overrightarrow{\omega }}$ is the geometrical rotation.

\subsection{First Consequences of a Unit of Action}

In the remainder, we assume that there exists some natural unit of action,
namely the Planck constant $h$ (or the reduced value $\hslash =h/2\pi $),
and we are interested in the consequences of this hypothesis for the objects
defined on $\mathcal{H}$.

\subsubsection{New physical symmetry: The Symplectic Transform}

Taking into account the data of $\hslash $, we can define a set of linear
operators $\left\{ \mathbf{K}_{S}(\alpha )\right\} _{\alpha \in \mathbf{R}}$
(symplectic involutions) depending on the undimensional real parameter $%
\alpha $ by: 
\begin{equation}
<\overrightarrow{p},\overrightarrow{q}|\mathbf{K}_{S}(\alpha )|%
\overrightarrow{p_{1}},\overrightarrow{q_{1}}>=(\alpha /h)^{3}\exp [(i\alpha
/\hslash )(\overrightarrow{q}.\overrightarrow{p_{1}}-\overrightarrow{p}.%
\overrightarrow{q_{1}})]
\end{equation}

A simple checking proves that: 
\begin{equation}
\mathbf{K}_{S}^{\dagger }(\alpha )=\mathbf{K}_{S}(\alpha )\text{ and }%
\mathbf{K}_{S}(\alpha )^{2}=\mathbf{1}_{\mathcal{H}}
\end{equation}

So each $\mathbf{K}_{S}(\alpha )$ is a self-adjoint unitary operator, that
is an involution; and $\mathbf{K}_{S}(\alpha )$ depends on the symplectic
product $\overrightarrow{q}.\overrightarrow{p_{1}}-\overrightarrow{p}.%
\overrightarrow{q_{1}}$.

On the other hand, we can also define the unitary gauge transforms $\mathbf{U%
}_{G}(\xi )$ depending on the undimensional real parameter $\xi $ as: 
\begin{equation}
\mathbf{U}_{G}(\xi )\text{ }|\overrightarrow{p},\overrightarrow{q}>=\exp
[-(i\xi /\hslash )\overrightarrow{p}.\overrightarrow{q}]\text{ }|%
\overrightarrow{p},\overrightarrow{q}>
\end{equation}

Or: 
\begin{equation}
\mathbf{U}_{G}(\xi )=\exp [-(i\xi /\hslash )\overrightarrow{\mathbf{p}}.%
\overrightarrow{\mathbf{q}}]
\end{equation}

These operators $\mathbf{K}_{S}(\alpha )$ and $\mathbf{U}_{G}(\xi )$ are the
simplest ''new symmetries'' induced by the existence of $\hslash $.\newline
Combining $\mathbf{K}_{S}(\alpha )$ and $\mathbf{U}_{G}(\xi )$, we extend
the set $\left\{ \mathbf{K}_{S}(\alpha )\right\} _{\alpha \in \mathbf{R}}$
of involutions (unitary self-adjoint operators) to $\left\{ \mathbf{K}
_{S}(\alpha ,\xi )\right\} _{(\alpha ,\xi )\in \mathbf{R}^{2}}$ with: 
\begin{equation}
\mathbf{K}_{S}(\alpha ,\xi )=\mathbf{U}_{G}(\xi )^{\dagger }\mathbf{K}
_{S}(\alpha )\mathbf{U}_{G}(\xi )
\end{equation}

Now, each $\mathbf{K}_{S}(\alpha ,\xi )$ is mathematically a new symmetry.
But is it really possible that different non-equivalent $\mathbf{K}
_{S}(\alpha ,\xi )$ physically exist as a same time?

In fact, we see that modifying $\alpha $ and $\xi $ corresponds to a scaling
on $\hslash $. If we say that there exists a \textbf{unique} unit of action,
only one of these $\mathbf{K}_{S}(\alpha ,\xi )$ must be taken as
fundamental. Of course, this is not sufficient to specify the value of $%
\alpha $ and $\xi $; so we have to do a choice.

\textbf{Conclusion}\newline
We postulate in all the following, that \textbf{the} \textbf{fundamental
symplectic transform} is $\mathbf{K}_{S}$ defined as: 
\begin{equation}
\mathbf{K}_{S}=\mathbf{K}_{S}(1/2,1/2)
\end{equation}

Or:

\begin{equation}
<\overrightarrow{p},\overrightarrow{q}|\mathbf{K}_{S}|\overrightarrow{p_{1}},%
\overrightarrow{q_{1}}>=(1/2h)^{3}\exp [(i/2\hslash )(\overrightarrow{q}-%
\overrightarrow{q_{1}}).(\overrightarrow{p}+\overrightarrow{p_{1}})]
\label{symplectictransform}
\end{equation}
(Of course this choice contains a part of arbitrary due to possible others
unitary equivalent possibilities).

Now, $\mathbf{K}_{S}$ must be seen as a new physical symmetry (unclassical)
induced by the existence of ''$h$'' and then it must be added to Galileo
Group.

\subsubsection{''Passive'' and ''Active'' Representations of Observables}

We have seen in paragraph IV that we can mathematically associate to each
classical observable $f$, two self-adjoint operators $\mathbf{f}$ and $%
\mathbf{X}_{f}$ , $\mathbf{X}_{f}$ being homogeneous to $f$ divided by an
action.

So $\mathbf{f}$ and $\hslash \mathbf{X}_{f}$ are now two possible
representations of $f$ with the same physical homogeneity, but $\hslash 
\mathbf{X}_{f}$ is a non-classical observable. Moreover $\mathbf{f}$ and $%
\hslash \mathbf{X}_{f}$ are independent since $[\mathbf{f},\hslash \mathbf{X}
_{f}]=0$. Then, introducing $\hslash $ generates some apparent mathematical
ambiguity into the representation of observables: each observable can be
defined as a data or as a generator of a one-parameter group.

We will see in the remainder that our procedure of quantization allows to
remove this ambiguity.

\subsection{New Representation of Galileo Group}

Now, we consider $\mathbf{K}_{S}$ as a new physical symmetry, but without
any classical equivalent. So it is natural to assume that $\mathbf{K}_{S}$
does not interfere with the unitary operators $\mathbf{U}_{\beta }$ that
represent Galileo group, in other words $[\mathbf{K}_{S},\mathbf{U}_{\beta
}]=0$. This means that the (self-adjoint) generators $\mathbf{Y}\alpha $ of
Galileo group must verify $[\mathbf{K}_{S},\mathbf{Y}\alpha ]=0$. Since $%
\mathbf{K}_{S}$ is an involution, this last requirement can be written as: 
\begin{equation}
\mathbf{K}_{S}\mathbf{Y}\alpha \mathbf{K}_{S}=\mathbf{Y}\alpha \text{ or }%
\mathbf{Y}\alpha =\frac{1}{2}(\mathbf{Y}\alpha +\mathbf{K}_{S}\mathbf{Y}
\alpha \mathbf{K}_{S})
\end{equation}

Now, as seen in VII-A, the classical generators of Galileo group are $%
\mathbf{X}_{\overrightarrow{p}}=-i\overrightarrow{\nabla _{q}}$, $\mathbf{X}
_{\overrightarrow{q}}=i\overrightarrow{\nabla _{p}}$ and $\mathbf{X}_{%
\overrightarrow{q}\wedge \overrightarrow{p}}=(-i)(\overrightarrow{q}\wedge 
\overrightarrow{\nabla _{q}}+\overrightarrow{p}\wedge \overrightarrow{\nabla
_{p}})$. Using (\ref{symplectictransform}) and the fact that $\mathbf{K}_{S}$
is an involution, we find after a few algebra: 
\begin{equation}
\left\{ 
\begin{array}{l}
\mathbf{K}_{S}\mathbf{X}_{\overrightarrow{p}}\mathbf{K}_{S}=(1/2\hslash )(%
\overrightarrow{\mathbf{p}}+\mathbf{K}_{S}\overrightarrow{\mathbf{p}}\mathbf{%
\ K}_{S})=\mathbf{X}_{\overrightarrow{p}} \\ 
\mathbf{K}_{S}\mathbf{X}_{\overrightarrow{q}}\mathbf{K}_{S}=(1/2\hslash )(%
\overrightarrow{\mathbf{q}}-\mathbf{K}_{S}\overrightarrow{\mathbf{q}}\mathbf{%
\ K}_{S})=-\mathbf{X}_{\overrightarrow{q}} \\ 
\mathbf{K}_{S}\mathbf{X}_{\overrightarrow{q}\wedge \overrightarrow{p}}%
\mathbf{K}_{S}=\mathbf{X}_{\overrightarrow{q}\wedge \overrightarrow{p}}
\end{array}
\right.  \label{KSandclassgene}
\end{equation}

So, $\mathbf{X}_{\overrightarrow{p}}$ and $\mathbf{X}_{\overrightarrow{q}
\wedge \overrightarrow{p}}$ (generators of translation and rotations) are
effectively invariant under $\mathbf{K}_{S}$, but not $\mathbf{X}_{%
\overrightarrow{q}}$ (generator of Galileo boosts). So we must modify $%
\mathbf{X}_{\overrightarrow{q}}$ and we define the new generator $\mathbf{X}
_{\overrightarrow{q}}^{(1)}$ of Galileo boosts as: 
\begin{equation}
\mathbf{X}_{\overrightarrow{q}}^{(1)}=(1/2\hslash )(\overrightarrow{\mathbf{q%
}}+\mathbf{K}_{S}\overrightarrow{\mathbf{q}}\mathbf{K}_{S})=\hslash ^{-1}%
\overrightarrow{\mathbf{q}}+\mathbf{X}_{\overrightarrow{q}}
\end{equation}

Now we have: 
\begin{equation}
\mathbf{K}_{S}\mathbf{X}_{\overrightarrow{q}}^{(1)}\mathbf{K}_{S}=\mathbf{X}
_{\overrightarrow{q}}^{(1)}
\end{equation}

This new generator $\mathbf{X}_{\overrightarrow{q}}^{(1)}$ is invariant
under $\mathbf{K}_{S}$ and the new representation of Galileo boosts becomes: 
\begin{equation}
\exp \left[ i\overrightarrow{p_{0}}.\mathbf{X}_{\overrightarrow{q}
}^{(1)}\right] |\overrightarrow{p},\overrightarrow{q}>\text{ }=\exp
[(i/\hslash )\overrightarrow{p_{0}}.\overrightarrow{q}]\text{ }|%
\overrightarrow{p}+\overrightarrow{p_{0}},\overrightarrow{q}>
\end{equation}

As expected from general arguments of section VI-C, we find a supplementary
phase factor $\theta =(i/\hslash )\overrightarrow{p_{0}}.\overrightarrow{q}$
that modifies the classical representation, but only for Galileo boosts. The
representation of translations and rotations is unchanged.

To conclude, taking into account the remarks of the previous paragraph
VII-B2 on the representation of observables, we introduce the
''pre-quantum'' operators $\overrightarrow{\mathbf{P}_{*}}$, $%
\overrightarrow{\mathbf{Q}_{*}}$ and $\overrightarrow{\mathbf{J}_{*}}$,
homogeneous respectively to $\overrightarrow{p}$, $\overrightarrow{q}$ and $%
\overrightarrow{q}\wedge \overrightarrow{p}=\overrightarrow{l}$: 
\begin{equation}
\left\{ 
\begin{array}{l}
\overrightarrow{\mathbf{P}_{*}}=\hslash \mathbf{X}_{\overrightarrow{p}%
}=-i\hslash \overrightarrow{\nabla _{q}}=(1/2)(\overrightarrow{\mathbf{p}}+%
\mathbf{K}_{S}\overrightarrow{\mathbf{p}}\mathbf{K}_{S}) \\ 
\overrightarrow{\mathbf{Q}_{*}}=\hslash \mathbf{X}_{\overrightarrow{q}%
}^{(1)}=\overrightarrow{q}+i\hslash \overrightarrow{\nabla _{p}}=(1/2)(%
\overrightarrow{\mathbf{q}}+\mathbf{K}_{S}\overrightarrow{\mathbf{q}}\mathbf{%
K}_{S}) \\ 
\overrightarrow{\mathbf{J}_{*}}=\hslash \mathbf{X}_{\overrightarrow{l}%
}=-i\hslash \left( \overrightarrow{q}\wedge \overrightarrow{\nabla _{q}}+%
\overrightarrow{p}\wedge \overrightarrow{\nabla _{p}}\right)
\end{array}
\right.  \label{quantumgenerators}
\end{equation}

These operators are the generators of the new representation of Galileo
group through the equations: 
\begin{equation}
\left\{ 
\begin{array}{l}
\exp \left[ -(i/\hslash )\overrightarrow{q_{0}}.\overrightarrow{\mathbf{P}
_{*}}\right] |\overrightarrow{p},\overrightarrow{q}>=|\overrightarrow{p},%
\overrightarrow{q}+\overrightarrow{q_{0}}> \\ 
\exp \left[ (i/\hslash )\overrightarrow{p_{0}}.\overrightarrow{\mathbf{Q}_{*}%
}\right] |\overrightarrow{p},\overrightarrow{q}>=\exp [(i/\hslash )%
\overrightarrow{p_{0}}.\overrightarrow{q}]\text{ }|\overrightarrow{p}+%
\overrightarrow{p_{0}},\overrightarrow{q}> \\ 
\exp \left[ -(i/\hslash )\overrightarrow{\omega }.\overrightarrow{\mathbf{J}
_{*}}\right] |\overrightarrow{p},\overrightarrow{q}>=|\mathcal{R}_{%
\overrightarrow{\omega }}(\overrightarrow{q}),\mathcal{R}_{\overrightarrow{
\omega }}(\overrightarrow{p})>
\end{array}
\right.
\end{equation}
where $\mathcal{R}_{\overrightarrow{\omega }}$ is the geometrical rotation.

The index ''*'' on $\overrightarrow{\mathbf{P}_{*}}$, $\overrightarrow{%
\mathbf{Q}_{*}}$ and $\overrightarrow{\mathbf{J}_{*}}$ is used to
distinguish these ''pre-quantum operators'' from the true quantum ones that
will be obtained by our procedure of quantization.

Now, while the old generators $\mathbf{X}_{\overrightarrow{p}}$ and $\mathbf{%
\ X}_{\overrightarrow{q}}$ were commuting, the new ones $\overrightarrow{%
\mathbf{P}_{*}}$ and $\overrightarrow{\mathbf{Q}_{*}}$ verify: 
\begin{equation}
\lbrack \mathbf{P}_{*i},\mathbf{Q}_{*j}]=-i\hbar \delta _{ij}
\end{equation}

In fact, for any component of the operators $\overrightarrow{\mathbf{P}_{*}}$
, $\overrightarrow{\mathbf{Q}_{*}}$ and $\overrightarrow{\mathbf{J}_{*}}$,
the expression of the commutator $(i/\hslash )[\mathbf{A},\mathbf{B}]$
exactly corresponds to the Poisson Bracket of the associated classical
observables ($\overrightarrow{p}$, $\overrightarrow{q}$, $\overrightarrow{l}$
): 
\begin{equation}
\left\{ 
\begin{array}{l}
\lbrack \mathbf{P}_{*i},\mathbf{P}_{*j}]=[\mathbf{Q}_{*i},\mathbf{Q}_{*j}]=0
\\ 
\lbrack \mathbf{P}_{*i},\mathbf{Q}_{*j}]=-i\hbar \delta _{ij} \\ 
\lbrack \mathbf{J}_{*i},\mathbf{J}_{*j}]=i\hslash \epsilon _{ijk}\mathbf{J}
_{*k} \\ 
\lbrack \mathbf{J}_{*i},\mathbf{Q}_{*j}]=i\hslash \epsilon _{ijk}\mathbf{Q}
_{*k} \\ 
\lbrack \mathbf{J}_{*i},\mathbf{P}_{*j}]=i\hslash \epsilon _{ijk}\mathbf{P}
_{*k}
\end{array}
\right.  \label{pqjcommutators}
\end{equation}

So, independently of our specific procedure of quantization of observables
based on projection, it is natural to expect that these operators $%
\overrightarrow{\mathbf{P}_{*}}$, $\overrightarrow{\mathbf{Q}_{*}}$ and $%
\overrightarrow{\mathbf{J}_{*}}$ are the right representations of momentum,
position and angular momentum.

We will find that our procedure of quantization confirms this hypothesis.

\subsection{Angular Momentum and Spin}

Starting from $\overrightarrow{\mathbf{P}_{*}}$, $\overrightarrow{\mathbf{Q}
_{*}}$ and $\overrightarrow{\mathbf{J}_{*}}$, we define two new generators
that we call $\overrightarrow{\mathbf{L}_{*}}$ and $\overrightarrow{\mathbf{S%
}_{*}}$: 
\begin{equation}
\left\{ 
\begin{array}{l}
\overrightarrow{\mathbf{L}_{*}}=\overrightarrow{\mathbf{Q}_{*}}\wedge 
\overrightarrow{\mathbf{P}_{*}} \\ 
\overrightarrow{\mathbf{S}_{*}}=\overrightarrow{\mathbf{J}_{*}}-%
\overrightarrow{\mathbf{L}_{*}}=-i\hslash \overrightarrow{p}\wedge 
\overrightarrow{\nabla _{p}}-\hslash ^{2}\overrightarrow{\nabla _{p}}\wedge 
\overrightarrow{\nabla _{q}}
\end{array}
\right.
\end{equation}

$\overrightarrow{\mathbf{L}_{*}}$ and $\overrightarrow{\mathbf{S}_{*}}$ are
generators of rotations since their components verify $[\mathbf{A}_{i},%
\mathbf{A}_{j}]=i\hslash \epsilon _{ijk}\mathbf{A}_{k}$. But moreover we
have: 
\begin{equation}
\lbrack \mathbf{S}_{*i},\mathbf{Q}_{*j}]=[\mathbf{S}_{*i},\mathbf{P}_{*j}]=0
\end{equation}

So, anticipating on the remainder, $\overrightarrow{\mathbf{L}_{*}}$
generates orbital rotations, while $\overrightarrow{\mathbf{S}_{*}}$ defines
internal rotations, that is effects due to spin.

Nevertheless, since $\overrightarrow{\mathbf{S}_{*}}$ is built with true
vectors, it only generates integer spins.

This means that we cannot recover half-integer spins only starting with our
Configuration Space $\mathcal{H}=L^{2}(\mathcal{C})$. If we want to recover
all possible values of spin, we must add some external degrees of freedom to
our Hilbert space.

More precisely, we must assume that the basis of $\mathcal{H}$ is $\{|%
\overrightarrow{p},\overrightarrow{q}>\otimes $ $|\epsilon >\}$ with $%
\epsilon =0,\pm 1$. The value $\epsilon =0$ corresponds to the previous case
of integer spins, while the new values $\epsilon =\pm 1$ describe the
half-integer components. We must also modify the generator $\overrightarrow{
J_{*}}$ of rotations to take into account rotations of new degrees of
freedom: 
\begin{equation}
\overrightarrow{\mathbf{J}_{*}}=\hslash \mathbf{X}_{\overrightarrow{l}}+%
\frac{\hslash }{2}\overrightarrow{\Sigma }
\end{equation}
where $\overrightarrow{\Sigma }$ applied to $|\epsilon =0>$ gives $0$, and
applied to $|\epsilon =\pm 1>$, $\overrightarrow{\Sigma }$ reduces to Pauli
matrices.

Of course the operators $\overrightarrow{\mathbf{P}_{*}}$, $\overrightarrow{%
\mathbf{Q}_{*}}$ and $\overrightarrow{\mathbf{L}_{*}}$ are unchanged, and
only $\overrightarrow{\mathbf{S}_{*}}$ is modified.\bigskip

In the remainder we don't develop any further this extended framework,
because the heaviness of the formalism hides the ideas involved in our
procedure of quantization. So we continue to use our Hilbert space generated
by the basis $\{|\overrightarrow{p},\overrightarrow{q}>\}$, but now we know
that this frame can only generate integer spins, and then all the following
is devoted to bosonic particles.

\subsection{Discrete Symmetries}

\subsubsection{Parity}

Parity is defined as the linear operator $\mathbf{K}_{P}$ acting on states $|%
\overrightarrow{p},\overrightarrow{q}>$ as: 
\begin{equation}
\mathbf{K}_{P}|\overrightarrow{p},\overrightarrow{q}>=|-\overrightarrow{p},-%
\overrightarrow{q}>  \label{paritydef}
\end{equation}

\subsubsection{Time Reversal}

Classically speaking, Time Reversal transforms a state $(\overrightarrow{p},%
\overrightarrow{q})$ in $(-\overrightarrow{p},\overrightarrow{q})$. Then we
can try to represent Time Reversal as a linear operator acting on states $|%
\overrightarrow{p},\overrightarrow{q}>$, such that $\mathbf{K}_{T}|%
\overrightarrow{p},\overrightarrow{q}>$ $=|-\overrightarrow{p},%
\overrightarrow{q}>$. But if we take this definition, the linear condition
is uncompatible with Galileo boosts, because of the supplementary phase
factor introduced into the representation. To obtain a consistent result, we
must assume that $\mathbf{K}_{T}$ is an antilinear operator. So, we define $%
\mathbf{K}_{T}$ as an antilinear operator such that: 
\begin{equation}
\mathbf{K}_{T}|\overrightarrow{p},\overrightarrow{q}>\text{ }=|-%
\overrightarrow{p},\overrightarrow{q}>  \label{timereversaldef}
\end{equation}

\subsubsection{Symplectic Transform}

The symmetry $\mathbf{K}_{S}$ has been defined in paragraph VII-B1, and by
construction $\mathbf{K}_{S}$ commutes with all the previous symmetries.

\section{Irreducible Representation of Galileo Group}

Computations on the irreducible representations of Galileo group in phase
space have already been published (for example E. Prugove\v {c}ki\cite{prugo}
), but we need here to detail this analysis in order to prove the
consistency of our approach and also to introduce specific notations.

\subsection{Subspaces of irreducible representation of $P_{*}$ and $Q_{*}$}

We have seen that the generators $\overrightarrow{\mathbf{P}_{*}}$ and $%
\overrightarrow{\mathbf{Q}_{*}}$ verify $[\mathbf{P}_{*i},\mathbf{Q}
_{*j}]=-i\hbar \delta _{ij}$ and we know that irreducible representations of
these relations are obtained if $\overrightarrow{\mathbf{P}_{*}}\equiv
-i\hslash \overrightarrow{\nabla _{x}}$ and $\overrightarrow{\mathbf{Q}_{*}}
\equiv \overrightarrow{x}$ (for the moment $\overrightarrow{x}$ must be only
seen as a parameter). So we look for (unbounded) states ''$|\overrightarrow{x%
},\alpha >$'' such that: 
\begin{equation}
\left\{ 
\begin{array}{l}
<\overrightarrow{x},\alpha |\overrightarrow{\mathbf{P}_{*}}|\overrightarrow{p%
},\overrightarrow{q}>=-i\hslash \overrightarrow{\nabla _{x}}<\overrightarrow{
x},\alpha |\overrightarrow{p},\overrightarrow{q}> \\ 
<\overrightarrow{x},\alpha |\overrightarrow{\mathbf{Q}_{*}}|\overrightarrow{p%
},\overrightarrow{q}>=\overrightarrow{x}<\overrightarrow{x},\alpha |%
\overrightarrow{p},\overrightarrow{q}>
\end{array}
\right.
\end{equation}

If we call $\phi =<\overrightarrow{p},\overrightarrow{q}|\overrightarrow{x}
,\alpha >$, using the explicit expression of $\overrightarrow{\mathbf{P}_{*}}
$ and $\overrightarrow{\mathbf{Q}_{*}}$, we obtain: 
\begin{equation}
\left\{ 
\begin{array}{l}
\left( \overrightarrow{\nabla _{q}}+\overrightarrow{\nabla _{x}}\right) \phi
=0 \\ 
i\hslash \overrightarrow{\nabla _{p}}\phi =(\overrightarrow{x}-%
\overrightarrow{q})\phi
\end{array}
\right.
\end{equation}

The general solution of this system is: 
\begin{equation}
\phi =\Phi (\overrightarrow{x}-\overrightarrow{q})\exp [-(i/\hslash )%
\overrightarrow{p}.(\overrightarrow{x}-\overrightarrow{q})]
\end{equation}
where $\Phi $ is an arbitrary function.

So the states $\{|\overrightarrow{x},\alpha >\}$ solutions of the problem
are given by: 
\begin{equation}
<\overrightarrow{p},\overrightarrow{q}|\overrightarrow{x},\alpha >\text{ }
=\Phi _{\alpha }(\overrightarrow{x}-\overrightarrow{q})\exp [-(i/\hslash )%
\overrightarrow{p}.(\overrightarrow{x}-\overrightarrow{q})]  \label{statexn}
\end{equation}

Computing the scalar product $<\overrightarrow{y},\beta |\overrightarrow{x}
,\alpha >$ we obtain: 
\begin{equation}
\left\{ 
\begin{array}{l}
<\overrightarrow{y},\beta |\overrightarrow{x},\alpha >\text{ }=\text{ }
<<\beta |\alpha >>\delta (\overrightarrow{y}-\overrightarrow{x}) \\ 
<<\beta |\alpha >>\text{ }=h^{3}\int d^{3}\overrightarrow{q}\Phi _{\beta
}^{*}(\overrightarrow{q})\Phi _{\alpha }(\overrightarrow{q})
\end{array}
\right.  \label{reducedscalarp}
\end{equation}
where $h=2\pi \hslash $ is the Planck constant and $<<\beta |\alpha >>$
defines a reduced scalar product on the fields $\Phi _{\alpha }$.

Now, if we take a complete orthonormal basis $\{\Phi _{n}\}$ for the reduced
scalar product ($<<n|m>>=\delta _{n,m}$), the states $\{|\overrightarrow{x}
,n>\}$ defines a complete orthogonal basis for our Configuration Space and
we can define the bounded projectors $\Pi _{n}$ on each subspace generated
by the states $\{|\overrightarrow{x},n>\}_{\overrightarrow{x}\in \mathbf{R}
^{3}}$: 
\begin{equation}
\Pi _{n}=\int d^{3}\overrightarrow{x}\text{ }|\overrightarrow{x},n><%
\overrightarrow{x},n|
\end{equation}

Moreover, we have also the closure relation: 
\begin{equation}
\sum_{n}\Pi _{n}=\mathbf{1}_{\mathcal{H}}
\end{equation}

Each of these $\Pi _{n}$ defines a subspace of irreducible representation of
the commutation relations between $\overrightarrow{\mathbf{P}_{*}}$ and $%
\overrightarrow{\mathbf{Q}_{*}}$ and: 
\begin{equation}
\lbrack \overrightarrow{\mathbf{P}_{*}},\Pi _{n}]=[\overrightarrow{\mathbf{Q}
_{*}},\Pi _{n}]=0
\end{equation}

The action of space translations and Galileo boosts generated by $%
\overrightarrow{P_{*}}$ and $\overrightarrow{Q_{*}}$ on each subspace is
given by: 
\begin{equation}
\left\{ 
\begin{array}{l}
\exp \left[ -(i/\hslash )\overrightarrow{q_{0}}.\overrightarrow{\mathbf{P}
_{*}}\right] \text{ }|\overrightarrow{x},n>=|\overrightarrow{x}+%
\overrightarrow{q_{0}},n> \\ 
\exp \left[ (i/\hslash )\overrightarrow{q_{0}}.\overrightarrow{\mathbf{Q}_{*}%
}\right] \text{ }|\overrightarrow{x},n>=\exp [(i/\hslash )\overrightarrow{
p_{0}}.\overrightarrow{x}]\text{ }|\overrightarrow{x},n>
\end{array}
\right.
\end{equation}

Now, we look at irreducible representations of the rotation group, taking
into account the previous results.

\subsection{Irreducible representations of Rotations}

We have seen previously that the generator of rotations is $\overrightarrow{%
\mathbf{J}_{*}}=\overrightarrow{\mathbf{L}_{*}}+\overrightarrow{\mathbf{S}
_{*}}$ with $[\overrightarrow{\mathbf{L}_{*}},\overrightarrow{\mathbf{S}_{*}}
]=0$. Then: 
\begin{equation}
\exp \left[ -(i/\hslash )\overrightarrow{\omega }.\overrightarrow{\mathbf{J}
_{*}}\right] =\exp \left[ -(i/\hslash )\overrightarrow{\omega }.%
\overrightarrow{\mathbf{L}_{*}}\right] \exp \left[ -(i/\hslash )%
\overrightarrow{\omega }.\overrightarrow{\mathbf{S}_{*}}\right]
\end{equation}

Since by construction $\overrightarrow{\mathbf{L}_{*}}=\overrightarrow{%
\mathbf{Q}_{*}}\wedge \overrightarrow{\mathbf{P}_{*}}$ acts on each subspace 
$\Pi _{n}$ as the usual quantum operator of orbital angular momentum, we
have: 
\begin{equation}
\exp \left[ -(i/\hslash )\overrightarrow{\omega }.\overrightarrow{\mathbf{J}
_{*}}\right] \text{ }|\overrightarrow{x},n>\text{ }=\exp \left[ -(i/\hslash )%
\overrightarrow{\omega }.\overrightarrow{\mathbf{S}_{*}}\right] \text{ }|%
\mathcal{R}_{\overrightarrow{\omega }}(\overrightarrow{x}),n>
\label{prerotation}
\end{equation}

Then irreducible representations of rotations are only dependant of the
generators $\overrightarrow{\mathbf{S}_{*}}$. Moreover we know that these
representations are obtained by states $|S,m_{S}>$ such that $%
\overrightarrow{\mathbf{S}_{*}}^{2}|S,m_{S}>=\hslash ^{2}S(S+1)$ and $%
\mathbf{S}_{*z}|S,m_{S}>=\hslash m_{S}|S,m_{S}>$. Then we look for states $|%
\overrightarrow{x},S,m_{S}>$ solving the problem.

The action of $\overrightarrow{\mathbf{S}_{*}}$ on $|\overrightarrow{x},n>$
is given by: 
\begin{equation}
<\overrightarrow{p},\overrightarrow{q}|\overrightarrow{\mathbf{S}_{*}}|%
\overrightarrow{x},n>\text{ }=i\hslash \overrightarrow{\nabla _{p}}\wedge (%
\overrightarrow{p}+i\hslash \overrightarrow{\nabla _{q}})<\overrightarrow{p},%
\overrightarrow{q}|\overrightarrow{x},n>
\end{equation}

Taking the explicit expression (\ref{statexn}) of $<\overrightarrow{p},%
\overrightarrow{q}|\overrightarrow{x},n>$ we obtain: 
\begin{equation}
<\overrightarrow{p},\overrightarrow{q}|\overrightarrow{\mathbf{S}_{*}}|%
\overrightarrow{x},n>=-i\hslash (\overrightarrow{x}-\overrightarrow{q}
)\wedge \overrightarrow{\nabla }\Phi _{n}(\overrightarrow{x}-\overrightarrow{
q})\exp [-(i/\hslash )\overrightarrow{p}.(\overrightarrow{x}-\overrightarrow{
q})]
\end{equation}

So $\overrightarrow{\mathbf{S}_{*}}$ only acts on the field $\Phi _{n}$ as
the usual quantum orbital angular operator. Then we know that irreducible
representations are obtained by spherical harmonics and the states $|%
\overrightarrow{x},S,m_{S}>$ are defined by: 
\begin{equation}
\left\{ 
\begin{array}{l}
<\overrightarrow{p},\overrightarrow{q}|\overrightarrow{x},S,m_{S}>\text{ }
=\Phi _{S,m_{S}}(\overrightarrow{x}-\overrightarrow{q})\exp [-(i/\hslash )%
\overrightarrow{p}.(\overrightarrow{x}-\overrightarrow{q})] \\ 
\Phi _{S,m_{S}}(\overrightarrow{x})=\Psi (||\overrightarrow{x}
||)Y_{S}^{m_{S}}(\overrightarrow{x}/||\overrightarrow{x}||)
\end{array}
\right.  \label{irreduciblestates}
\end{equation}

$\Psi $ is an arbitrary function normalized with the reduced scalar product
defined in (\ref{reducedscalarp}): 
\begin{equation}
\int_{0}^{\infty }x^{2}dx\text{ }|\Psi (x)|^{2}=h^{-3}  \label{normalization}
\end{equation}

Now the action of the rotation group on the orthogonal basis $|%
\overrightarrow{x},S,m_{S}>$ can be deduced from (\ref{prerotation}): 
\begin{equation}
\exp \left[ -(i/\hslash )\overrightarrow{\omega }.\overrightarrow{\mathbf{J}
_{*}}\right] \text{ }|\overrightarrow{x},S,m>\text{ } =%
\sum_{m_{1}}R_{m_{1},m}^{S}(\overrightarrow{\omega })\text{ }|\mathcal{R}_{%
\overrightarrow{\omega }}(\overrightarrow{x}),S,m_{1}>
\end{equation}
where $R_{m_{1},m}^{S}(\overrightarrow{\omega })$ is the irreducible matrix
of rotation.

\subsection{Conclusion}

The final result is that irreducible subspaces of the Galileo group are
given by the projectors $\Pi _{S}$ on the states generated by the orthogonal
basis $\{|\overrightarrow{x},S,m>\}$, that is: 
\begin{equation}
\Pi _{S}=\sum_{m}\int d^{3}\overrightarrow{x}\text{ }|\overrightarrow{x}
,S,m><\overrightarrow{x},S,m|
\end{equation}

Each of these subspaces depends of course on the value of $S$ (integer), but
also depends on an arbitrary function $\Psi $ normalized by (\ref
{normalization}).

\subsection{Action of discrete symmetries on an irreducible subspace}

\subsubsection{Parity}

Following the definition (\ref{paritydef}) of the operator $\mathbf{K}_{P}$
we find: 
\begin{equation}
<\overrightarrow{p},\overrightarrow{q}|\mathbf{K}_{P}|\overrightarrow{x}
,S,m_{S}>=\Phi _{S,m_{S}}(\overrightarrow{x}+\overrightarrow{q})\exp
[(i/\hslash )\overrightarrow{p}.(\overrightarrow{x}+\overrightarrow{q})]
\end{equation}

Now, using the parity of the spherical harmonics, we find: 
\begin{equation}
\mathbf{K}_{P}\text{ }|\overrightarrow{x},S,m_{S}>\text{ }=(-1)^{S}|-%
\overrightarrow{x},S,m_{S}>
\end{equation}

\subsubsection{Time Reversal}

Following the definition (\ref{timereversaldef}) of the antilinear operator $%
\mathbf{K}_{T}$ we find: 
\begin{equation}
<\overrightarrow{p},\overrightarrow{q}|\mathbf{K}_{T}|\overrightarrow{x}
,S,m_{S}>=\Phi _{S,m_{S}}^{*}(\overrightarrow{x}-\overrightarrow{q})\exp
[-(i/\hslash )\overrightarrow{p}.(\overrightarrow{x}-\overrightarrow{q})]
\end{equation}

Using the definition of $\Phi _{S,m_{S}}$ and the relation between spherical
harmonic and its conjugate, we find: 
\begin{equation}
<\overrightarrow{p},\overrightarrow{q}|\mathbf{K}_{T}|\overrightarrow{x}
,S,m>=(-1)^{m}\Psi ^{*}(|\overrightarrow{x}-\overrightarrow{q}
|)Y_{S}^{-m}\exp [-(i/\hslash )\overrightarrow{p}.(\overrightarrow{x}-%
\overrightarrow{q})]
\end{equation}

We deduce that each subspace $\Pi _{S}$ is invariant by Time Reversal, only
if the unknown function $\Psi $ verifies $\Psi ^{*}=\alpha \Psi $.

Since $\Psi $\ is always defined up to a constant phase factor\textbf{, we
assume in the remainder that }$\Psi $\textbf{\ is real.}

Then $\mathbf{K}_{T}$ is an antilinear operator that verifies: 
\begin{equation}
\mathbf{K}_{T}|\overrightarrow{x},S,m_{S}>\text{ }=(-1)^{m_{S}}|%
\overrightarrow{x},S,-m_{S}>
\end{equation}

\subsubsection{Symplectic Transform}

Using the definition (\ref{symplectictransform}) of the operator $\mathbf{K}
_{S}$ we find: 
\begin{equation}
<\overrightarrow{p},\overrightarrow{q}|\mathbf{K}_{S}|\overrightarrow{x}
,S,m_{S}>=\Phi _{S,m_{S}}(\overrightarrow{q}-\overrightarrow{x})\exp
[-(i/\hslash )\overrightarrow{p}.(\overrightarrow{x}-\overrightarrow{q})]
\end{equation}

Because of the parity of spherical harmonics, we find: 
\begin{equation}
\mathbf{K}_{S}|\overrightarrow{x},S,m_{S}>=(-1)^{S}|\overrightarrow{x}
,S,m_{S}>  \label{KSaction}
\end{equation}

Then the non-classical symmetry $\mathbf{K}_{S}$ acts on $|\overrightarrow{x}
,S,m_{S}>$ as an intrinsic parity.\medskip

We conclude that the discrete symmetries $\mathbf{K}_{P}$, $\mathbf{K}_{T}$,
and $\mathbf{K}_{S}$ are represented on each subspace $\Pi _{S}$, if we
assume the unknown function $\Psi $ to be real.

\section{Axiom of Quantization}

Now any subspace $\mathcal{H}_{S}=Ran(\Pi _{S})$ corresponds to our
requirements defined in section VI-C and a quantization is the projection
from the global Hilbert space $\mathcal{H}$ on one of these subspaces $%
\mathcal{H}_{S}$.

As expected, the action of all symmetries on the states ''$|\overrightarrow{x%
},S,m_{S}>$'' correspond exactly with the quantum definition. But for the
moment the quantities $\overrightarrow{x},S,m_{S}$ are only mathematical
parameters and are not related to physical observables (even if we can guess
their meaning).

On the other hand, each $\mathcal{H}_{S}$ is defined by the integer
parameter $S$ and the real function $\Psi (x)$. The essential effect of this
function is to introduce a specific length scale $\lambda $ as a
characteristic of the representation. To make explicit this dependence, we
reduce $\Psi (x)$ to a purely mathematical (undimensional) function $\Psi
_{0}$ by the scaling: 
\begin{equation}
\Psi (x)=(\lambda h)^{-3/2}\Psi _{0}(x/\lambda )  \label{undimensionalpsi}
\end{equation}

The condition of normalization (\ref{normalization}) becomes: 
\begin{equation}
\int_{0}^{\infty }u^{2}du\Psi _{0}(u)^{2}=1
\end{equation}

To end, if we look at the main physical consequences of our procedure of
quantization, we see that the important result is that a particle associated
with a subspace $\mathcal{H}_{S}$ possesses an internal structure. This
structure is defined by the intrinsic properties for rotations (spin $S$),
and by a specific length scale $\lambda $. Then our procedure generates
naturally spins for particle, but on the other hand, we also find that a
particle must be associated with a natural length scale.

At this stage, if we want to give a physical meaning to $\lambda $, we must
specify what we mean by ''particles''. Of course, elementary objects such as
electrons or protons are particles, but atoms with frozen internal degrees
of freedom can be also seen as particles.

\textbf{The case of true elementary particles}

In non-relativistic quantum mechanics, we cannot build a natural length unit 
$\lambda $ for a true particle, but in relativistic quantum mechanics the
Compton wave length $\lambda _{c}=h/Mc$ defines such intrinsic length scale.
So it is natural to look at $\lambda $ as the preceding of the relativistic
quantity $\lambda _{c}$. This means that $\lambda $ must be always a very
small quantity (in regard to all other classical length scales).

\textbf{The case of complex particles (atoms)}

First, we must assume that all internal degrees of freedom are frozen, to be
able to describe the system only using external dynamical quantities. In
this case $\lambda $ simply represents the geometrical size of the system.
Moreover, we must always assume that $\lambda $ is very small in regard to
all other classical length scales of the problem, because the existence of
some length scale of order of $\lambda $ implies a dynamical effect on the
internal degrees of freedom of the system.

\section{The Basic Quantized Observables}

In all the remainder the subspace $\mathcal{H}_{S}$ is assumed to be fixed,
and following always our procedure of section VI-C, we can look how
classical observables are transformed through quantization.

We recall that, starting from a classical quantity $f(\overrightarrow{p},%
\overrightarrow{q})$, we build first the associated classical operator $%
\mathbf{f}$, and then we quantify $\mathbf{f}$ by taking the projection $\Pi
_{S}.\mathbf{f}.\Pi _{S}$.

\subsection{Quantum Operators of Position and Momentum}

So, we define the quantum operator of position $\overrightarrow{\mathbf{Q}}$
and momentum $\overrightarrow{\mathbf{P}}$ as: 
\begin{equation}
\left\{ 
\begin{array}{l}
\overrightarrow{\mathbf{Q}}=\Pi _{S}\overrightarrow{\mathbf{q}}\Pi _{S} \\ 
\overrightarrow{\mathbf{P}}=\Pi _{S}\overrightarrow{\mathbf{p}}\Pi _{S}
\end{array}
\right.
\end{equation}

First, we remark that the action of the symplectic transform $K_{S}$ on the
states $|\overrightarrow{x},S,m_{S}>$ (\ref{KSaction}) implies: 
\begin{equation}
K_{S}\Pi _{S}=\Pi _{S}K_{S}=(-1)^{S}\Pi _{S}  \label{KSprojected}
\end{equation}

Now, we have the definitions (\ref{quantumgenerators}) of $\overrightarrow{
P_{*}}$ and $\overrightarrow{Q_{*}}$: 
\begin{equation}
\left\{ 
\begin{array}{l}
\overrightarrow{P_{*}}=(1/2)(\overrightarrow{\mathbf{p}}+K_{S}%
\overrightarrow{\mathbf{p}}K_{S}) \\ 
\overrightarrow{Q_{*}}=(1/2)(\overrightarrow{\mathbf{q}}+K_{S}%
\overrightarrow{\mathbf{q}}K_{S})
\end{array}
\right.
\end{equation}

Projecting these relations with $\Pi _{S}$ and taking into account (\ref
{KSprojected}), we obtain: 
\begin{equation}
\left\{ 
\begin{array}{l}
\Pi _{S}\overrightarrow{P_{*}}=\Pi _{S}\overrightarrow{P_{*}}\Pi _{S}=\Pi
_{S}\overrightarrow{\mathbf{p}}\Pi _{S}=\overrightarrow{\mathbf{P}} \\ 
\Pi _{S}\overrightarrow{Q_{*}}=\Pi _{S}\overrightarrow{Q_{*}}\Pi _{S}=\Pi
_{S}\overrightarrow{\mathbf{q}}\Pi _{S}=\overrightarrow{\mathbf{Q}}
\end{array}
\right.  \label{pqquantified}
\end{equation}
(we recall that by construction $\Pi _{S}$ commutes with $\overrightarrow{
P_{*}}$ and $\overrightarrow{Q_{*}}$).

And then: 
\begin{equation}
\left\{ 
\begin{array}{l}
<\overrightarrow{x},S,m|\overrightarrow{\mathbf{P}}|\overrightarrow{y}
,S,m_{1}>=-i\hslash (\overrightarrow{\nabla }\delta )(\overrightarrow{x}-%
\overrightarrow{y})\delta _{m,m_{1}} \\ 
<\overrightarrow{x},S,m|\overrightarrow{\mathbf{Q}}|\overrightarrow{y}
,S,m_{1}>=\overrightarrow{x}\delta (\overrightarrow{x}-\overrightarrow{y}
)\delta _{m,m_{1}}
\end{array}
\right.
\end{equation}

\textbf{Conclusion}

We find that the quantized version $\overrightarrow{\mathbf{P}}$ and $%
\overrightarrow{\mathbf{Q}}$ of the classical observables $\overrightarrow{p}
$ and $\overrightarrow{q}$ correspond precisely with the usual quantum
operators.

Moreover, these operators are also the restriction of the generators of
translations and Galileo boosts. This means that, in the ''quantum world''
these observables are exactly the generators of symmetries (translation and
Galileo boosts) as in the classical picture through Poisson brackets.

We noticed that, as indicated previously, our procedure allows to remove the
ambiguity between ''passive and active'' representations of $\overrightarrow{
p}$ and $\overrightarrow{q}$: in fact they give the same operator.

But, we will see in the following paragraph that this result cannot be
extended to angular momentum because of spin: the projection of the
classical angular momentum $\overrightarrow{\mathbf{q}}\wedge 
\overrightarrow{\mathbf{p}}$ is not in general the restriction of the
generator of rotations $\overrightarrow{J_{*}}$. In this case we must use
other arguments to choose the quantum observable associated with angular
momentum.

\subsection{Quantum angular momentum and Spin}

We know from (\ref{quantumgenerators}) that the generator of rotations is $%
\overrightarrow{\mathbf{J}_{*}}=-i\hslash \left( \overrightarrow{q}\wedge 
\overrightarrow{\nabla _{q}}+\overrightarrow{p}\wedge \overrightarrow{\nabla
_{p}}\right) $ and by construction $\Pi _{S}$ commutes with $\overrightarrow{%
\mathbf{J}_{*}}$. Moreover we have seen in (\ref{pqjcommutators}) that the
commutators of $\overrightarrow{\mathbf{P}_{*}}$, $\overrightarrow{\mathbf{Q}
_{*}}$ and $\overrightarrow{\mathbf{J}_{*}}$ correspond exactly to the
expression of the Poisson Brackets of the classical quantities $%
\overrightarrow{p}$, $\overrightarrow{q}$ and $\overrightarrow{l}=%
\overrightarrow{q}\wedge \overrightarrow{p}$. Now, since we have proved that
the restriction $\overrightarrow{\mathbf{P}}$ and $\overrightarrow{\mathbf{Q}%
}$ of $\overrightarrow{\mathbf{P}_{*}}$ and $\overrightarrow{\mathbf{Q}_{*}}$
are the quantum operators of momentum and position, we can also guess that
the ''quantum angular momentum'' $\overrightarrow{\mathbf{J}}$ is the
restriction of the generator $\overrightarrow{\mathbf{J}_{*}}$. So we
define: 
\begin{equation}
\overrightarrow{\mathbf{J}}=\Pi _{S}\overrightarrow{\mathbf{J}_{*}}=%
\overrightarrow{\mathbf{J}_{*}}\Pi _{S}=\Pi _{S}\overrightarrow{\mathbf{J}
_{*}}\Pi _{S}
\end{equation}

Of course the general definitions of section III for observables attests
that $\overrightarrow{\mathbf{J}}$ (as a self adjoint operator commuting
with $\Pi _{S}$) is mathematically a possible observable. But up to now, we
have only defined quantum version of classical observables by our procedure
of quantization. So we must first find the relation connecting $%
\overrightarrow{\mathbf{J}}$ to the quantized version $\Pi _{S}%
\overrightarrow{\mathbf{l}}\Pi _{S}$ of the classical angular momentum $%
\overrightarrow{l}=\overrightarrow{q}\wedge \overrightarrow{p}$. The
following lines are devoted to this question.

Taking into account the explicit expression of $\overrightarrow{\mathbf{P}
_{*}}$ and $\overrightarrow{\mathbf{Q}_{*}}$, we rewrite $\overrightarrow{%
\mathbf{J}_{*}}$ as: 
\begin{equation}
\overrightarrow{\mathbf{J}_{*}}=\overrightarrow{\mathbf{q}}\wedge 
\overrightarrow{\mathbf{P}_{*}}+\overrightarrow{\mathbf{Q}_{*}}\wedge 
\overrightarrow{\mathbf{p}}-\overrightarrow{\mathbf{q}}\wedge 
\overrightarrow{\mathbf{p}}
\end{equation}

Using the equations (\ref{pqquantified}) related to $\overrightarrow{\mathbf{%
\ \ P}}$ and $\overrightarrow{\mathbf{Q}}$ we find: 
\begin{equation}
\overrightarrow{\mathbf{J}}=2\overrightarrow{\mathbf{Q}}\wedge 
\overrightarrow{\mathbf{P}}-\Pi _{S}\overrightarrow{\mathbf{q}}\wedge 
\overrightarrow{\mathbf{p}}\Pi _{S}  \label{Jexpression1}
\end{equation}

But we know also that the generator of rotations $\overrightarrow{\mathbf{J}
_{*}}$ can be divided into $\overrightarrow{\mathbf{J}_{*}}=\overrightarrow{%
\mathbf{L}_{*}}+\overrightarrow{\mathbf{S}_{*}}$ with $\overrightarrow{%
\mathbf{L}_{*}}=\overrightarrow{\mathbf{Q}_{*}}\wedge \overrightarrow{%
\mathbf{P}_{*}}$, and by construction $\Pi _{S}$ commutes with $%
\overrightarrow{\mathbf{L}_{*}}$ and $\overrightarrow{\mathbf{S}_{*}}$. So
we can define two new quantum observables $\overrightarrow{\mathbf{L}}$ and $%
\overrightarrow{\mathbf{S}}$ as the restriction of $\overrightarrow{\mathbf{L%
}_{*}}$ and $\overrightarrow{\mathbf{S}_{*}}$: 
\begin{equation}
\left\{ 
\begin{array}{l}
\overrightarrow{\mathbf{L}}=\Pi _{S}\overrightarrow{\mathbf{L}_{*}}=%
\overrightarrow{\mathbf{L}_{*}}\Pi _{S}=\overrightarrow{\mathbf{Q}}\wedge 
\overrightarrow{\mathbf{P}} \\ 
\overrightarrow{\mathbf{S}}=\Pi _{S}\overrightarrow{\mathbf{S}_{*}}=%
\overrightarrow{\mathbf{S}_{*}}\Pi _{S} \\ 
\overrightarrow{\mathbf{J}}=\overrightarrow{\mathbf{L}}+\overrightarrow{%
\mathbf{S}}
\end{array}
\right.  \label{Jexpression2}
\end{equation}

We call $\overrightarrow{\mathbf{L}}$ ''orbital angular momentum'' and $%
\overrightarrow{\mathbf{S}}$ ''spin momentum''.

Now using equations (\ref{Jexpression1}) and (\ref{Jexpression2}) we find
finally: 
\begin{equation}
\Pi _{S}\overrightarrow{\mathbf{q}}\wedge \overrightarrow{\mathbf{p}}\Pi
_{S}=\overrightarrow{\mathbf{L}}-\overrightarrow{\mathbf{S}}=\overrightarrow{%
\mathbf{Q}}\wedge \overrightarrow{\mathbf{P}}-\overrightarrow{\mathbf{S}}
\end{equation}

This shows that the quantized version of the classical angular momentum $%
\overrightarrow{l}=\overrightarrow{q}\wedge \overrightarrow{p}$ does not
correspond in general with the generator $\overrightarrow{\mathbf{J}}$ (or $%
\overrightarrow{\mathbf{L}}$).

In fact we find that $\Pi _{S}\overrightarrow{\mathbf{q}}\wedge 
\overrightarrow{\mathbf{p}}\Pi _{S}=\overrightarrow{\mathbf{L}}=%
\overrightarrow{\mathbf{J}}$ only in the case of scalar particles ($S=0$).

Then, for $S=0$, it is true that the quantized version of the classical
angular momentum $\overrightarrow{l}=\overrightarrow{q}\wedge 
\overrightarrow{p}$ is the generator $\overrightarrow{\mathbf{J}}=%
\overrightarrow{\mathbf{L}}=\overrightarrow{\mathbf{Q}}\wedge 
\overrightarrow{\mathbf{P}}$ of rotations. This is because we call $%
\overrightarrow{\mathbf{L}}$ ''orbital angular momentum''. But this
correspondence fails for $S\neq 0$, because of spin variables.

This result is very simple to understand, if we remind that $\overrightarrow{%
\mathbf{S}}$ cannot be obtained as the quantization of any classical
observable. So, introducing spins in our formalism is equivalent to define
non-classical observables. Then each subspace $\Pi _{S}$, for $S\neq 0$,
cannot be described only using quantized version of classical observables,
and this generates the non-equivalence between $\Pi _{S}\overrightarrow{%
\mathbf{l}}\Pi _{S}$ and $\overrightarrow{\mathbf{J}}$ (or $\overrightarrow{%
\mathbf{L}}$).

Nevertheless, if we look at the case $S\neq 0$ as an extension of $S=0$, we 
\textbf{must postulate} that $\overrightarrow{\mathbf{J}}$ is the physical
quantum observable of angular momentum, precisely because $\overrightarrow{%
\mathbf{J}}$ is always the generator of rotations. Moreover, $%
\overrightarrow{\mathbf{L}}$ and $\overrightarrow{\mathbf{S}}$ are also
quantum observables as the generator of orbital rotations and as the
generator of spin rotations.

To end, the parameter $S$ and the non-classical observable $\overrightarrow{%
\mathbf{S}}$ only appear because we look for irreducible representations of
Galileo group, in order to apply our general procedure of quantization. Then
we recover the usual result that spins are ''purely quantum
objects''.\bigskip

\textbf{Conclusion}\newline
We summarize the previous discussion in two points:

\begin{itemize}
\item  In the case of scalar particles ($S=0$), it is true that the
generator of rotations $\overrightarrow{\mathbf{J}}=\overrightarrow{\mathbf{L%
}}=\overrightarrow{\mathbf{Q}}\wedge \overrightarrow{\mathbf{P}}$ is the
quantized version of the classical angular momentum $\overrightarrow{l}=%
\overrightarrow{q}\wedge \overrightarrow{p}$. This is because we call $%
\overrightarrow{\mathbf{L}}$ ''orbital angular momentum''.

\item  In the case of non-scalar particle ($S\neq 0$), it is false that the
generator of rotations $\overrightarrow{\mathbf{J}}=\overrightarrow{\mathbf{L%
}}+\overrightarrow{\mathbf{S}}=\overrightarrow{\mathbf{Q}}\wedge 
\overrightarrow{\mathbf{P}}+\overrightarrow{\mathbf{S}}$ is the quantized
version of $\overrightarrow{l}=\overrightarrow{q}\wedge \overrightarrow{p}$.
We must postulate that $\overrightarrow{\mathbf{J}}$ is the physical
observable of angular momentum precisely because $\overrightarrow{\mathbf{J}}
$ is always the generator of rotations. The operators $\overrightarrow{%
\mathbf{L}}$ and $\overrightarrow{\mathbf{S}}$ are also quantum observables
as generators of orbital rotations and spin rotations (in fact $%
\overrightarrow{\mathbf{L}}=\Pi _{S}\overrightarrow{\mathbf{q}}\Pi %
_{S}\wedge \Pi _{S}\overrightarrow{\mathbf{p}}\Pi _{S}\neq \Pi _{S}%
\overrightarrow{\mathbf{q}}\wedge \overrightarrow{\mathbf{p}}\Pi _{S}$).
\end{itemize}

\subsection{Semi-Classical States and Semi-Classical Events}

\subsubsection{Semi-Classical States}

In the global Hilbert space $\mathcal{H}$, we have the closure relation: 
\begin{equation}
\mathbf{1}_{\mathcal{H}}=\int d^{3}\overrightarrow{p}d^{3}\overrightarrow{q}%
\text{ }|\overrightarrow{p},\overrightarrow{q}><\overrightarrow{p},%
\overrightarrow{q}|
\end{equation}

Then: 
\begin{equation}
\Pi _{S}=\int d^{3}\overrightarrow{p}d^{3}\overrightarrow{q}\text{ }\Pi _{S}|%
\overrightarrow{p},\overrightarrow{q}><\overrightarrow{p},\overrightarrow{q}
|\Pi _{S}  \label{quantumclosure}
\end{equation}

We deduce that the states $|\xi _{\overrightarrow{p},\overrightarrow{q}%
},S>=\Pi _{S}|\overrightarrow{p},\overrightarrow{q}>$ define an overcomplete
basis of our physical Hilbert space $\mathcal{H}_{S}$ (cf section VI-B3).

Moreover using (\ref{irreduciblestates}): 
\begin{equation}
<\xi _{\overrightarrow{p},\overrightarrow{q}},S|\xi _{\overrightarrow{p},%
\overrightarrow{q}},S>=\sum_{m}\int d^{3}\overrightarrow{x}\text{ }|\Phi
_{S,m}(\overrightarrow{x})|^{2}=(2S+1)h^{-3}
\end{equation}

In fact these states are coherent states for Galileo group and have a lot of
applications in standard Quantum Dynamics (E. Prugove\v {c}ki\cite{prugo} ,
R. Omn\`{e}s\cite{Omnes} , J.R. Perelomov or J.R. Klauder\cite{Klauder} ).

\subsubsection{Semi-Classical Configuration Events}

If we take a classical event ''$A$'' in Phase Space, it is represented in $%
\mathcal{H}$ by the projector $\Pi (A)$: 
\begin{equation}
\Pi (A)=\int_{A}d^{3}\overrightarrow{p}d^{3}\overrightarrow{q}\text{ }|%
\overrightarrow{p},\overrightarrow{q}><\overrightarrow{p},\overrightarrow{q}|
\end{equation}

Quantifying $\Pi (A)$, we obtain $\Pi _{S}(A)$: 
\begin{equation}
\Pi _{S}(A)=\Pi _{S}\Pi (A)\Pi _{S}=\int_{A}d^{3}\overrightarrow{p}d^{3}%
\overrightarrow{q}\text{ }|\xi _{\overrightarrow{p},\overrightarrow{q}
},S><\xi _{\overrightarrow{p},\overrightarrow{q}},S|
\end{equation}

As indicated in section VI-B3, $\Pi _{S}(A)$ is not any more a true
projector, and then $\Pi _{S}(A)$ must be seen as a ''fuzzy'' quantum event
(quasiprojector), or a semi-classical event (R. Omn\`{e}s\cite{Omnes}).

Now taking the trace of $\Pi _{S}(A)$ we obtain: 
\begin{equation}
Tr(\Pi _{S}(A))=(2S+1)\mathcal{V}(A)h^{-3}
\end{equation}
where $\mathcal{V}(A)$ is the volume of ''$A$'' in Phase Space.

Then $Tr(\Pi _{S}(A))$ gives exactly the semi-classical number of quantum
states contained in the volume ''$A$'' of Phase Space. This shows that $\Pi
_{S}(A)$ is really a physical intermediate between classical and quantum
world.

\subsection{Semi-Classical States and Quantum Observables}

\subsubsection{Expectation values}

For $\overrightarrow{\mathbf{P}}$, $\overrightarrow{\mathbf{Q}}$, $%
\overrightarrow{\mathbf{L}}=\overrightarrow{\mathbf{Q}}\wedge 
\overrightarrow{\mathbf{P}}$ and $\overrightarrow{\mathbf{S}}$, we find
after a few algebra: 
\begin{equation}
\left\{ 
\begin{array}{l}
<\xi _{\overrightarrow{p},\overrightarrow{q}},S|\overrightarrow{\mathbf{P}}
|\xi _{\overrightarrow{p},\overrightarrow{q}},S>=\mathcal{N}\overrightarrow{p%
} \\ 
<\xi _{\overrightarrow{p},\overrightarrow{q}},S|\overrightarrow{\mathbf{Q}}
|\xi _{\overrightarrow{p},\overrightarrow{q}},S>=\mathcal{N}\overrightarrow{q%
} \\ 
<\xi _{\overrightarrow{p},\overrightarrow{q}},S|\overrightarrow{\mathbf{L}}
|\xi _{\overrightarrow{p},\overrightarrow{q}},S>=\mathcal{N}\overrightarrow{q%
}\wedge \overrightarrow{p} \\ 
<\xi _{\overrightarrow{p},\overrightarrow{q}},S|\overrightarrow{\mathbf{S}}
|\xi _{\overrightarrow{p},\overrightarrow{q}},S>=0
\end{array}
\right.
\end{equation}
where $\mathcal{N}=<\xi _{\overrightarrow{p},\overrightarrow{q}},S|\xi _{%
\overrightarrow{p},\overrightarrow{q}},S>=(2S+1)h^{-3}$.

These results are precisely what we expect for semi-classical states.

\subsubsection{Splitting of observables on semi-classical states}

Moreover, since $\overrightarrow{\mathbf{P}}=\Pi _{S}\overrightarrow{\mathbf{%
\ \ p}}\Pi _{S}$ and $\overrightarrow{\mathbf{Q}}=\Pi _{S}\overrightarrow{%
\mathbf{q}}\Pi _{S}$ we have: 
\begin{equation}
\left\{ 
\begin{array}{l}
\overrightarrow{\mathbf{P}}=\int d^{3}\overrightarrow{p}d^{3}\overrightarrow{
q}\text{ }\overrightarrow{p}\text{ }|\xi _{\overrightarrow{p},%
\overrightarrow{q}},S><\xi _{\overrightarrow{p},\overrightarrow{q}},S| \\ 
\overrightarrow{\mathbf{Q}}=\int d^{3}\overrightarrow{p}d^{3}\overrightarrow{
q}\text{ }\overrightarrow{q}\text{ }|\xi _{\overrightarrow{p},%
\overrightarrow{q}},S><\xi _{\overrightarrow{p},\overrightarrow{q}},S|
\end{array}
\right.  \label{petqquantumsplitting}
\end{equation}

More generally, the quantized version $\mathbf{F}$ of any classical
observable $f(\overrightarrow{p},\overrightarrow{q})$ possesses the
splitting: 
\begin{equation}
\mathbf{F=}\Pi _{S}f(\overrightarrow{\mathbf{p}},\overrightarrow{\mathbf{q}})%
\Pi _{S}=\int d^{3}\overrightarrow{p}d^{3}\overrightarrow{q}\text{ }f(%
\overrightarrow{p},\overrightarrow{q})\text{ }|\xi _{\overrightarrow{p},%
\overrightarrow{q}},S><\xi _{\overrightarrow{p},\overrightarrow{q}},S|
\end{equation}

\subsection{Quantized Observables and Statistics}

If we interest to Statistics, the section VI-B shows that we must introduce
a density operator $\mathbf{D}$ with $\Pi _{S}\mathbf{D}=\mathbf{D}\Pi _{S}=%
\mathbf{D}$.

Moreover the expectation values of position and momentum are given by $<%
\overrightarrow{\mathbf{P}}>=Tr(\mathbf{D.}\overrightarrow{\mathbf{P}})$,
and $<\overrightarrow{\mathbf{Q}}>=Tr(\mathbf{D}.\overrightarrow{\mathbf{Q}}
) $.

But because $\overrightarrow{\mathbf{P}}=\Pi _{S}\overrightarrow{\mathbf{p}}
\Pi _{S}$, $\overrightarrow{\mathbf{Q}}=\Pi _{S}\overrightarrow{\mathbf{q}}
\Pi _{S}$ and $\mathbf{D}=\Pi _{S}\mathbf{D}\Pi _{S}$, we have also $<%
\overrightarrow{\mathbf{P}}>=Tr(\mathbf{D}.\overrightarrow{\mathbf{p}})$,
and $<\overrightarrow{\mathbf{Q}}>=Tr(\mathbf{D}.\overrightarrow{\mathbf{q}}
) $. So we can use our results of III-B3 on classical observables to obtain: 
\begin{equation}
\left\{ 
\begin{array}{l}
1=Tr(\mathbf{D})=\int d^{3}\overrightarrow{p}d^{3}\overrightarrow{q}\rho (%
\overrightarrow{p},\overrightarrow{q}) \\ 
<\overrightarrow{\mathbf{P}}>=\int d^{3}\overrightarrow{p}d^{3}%
\overrightarrow{q}\text{ }\overrightarrow{p}\rho (\overrightarrow{p},%
\overrightarrow{q}) \\ 
<\overrightarrow{\mathbf{Q}}>=\int d^{3}\overrightarrow{p}d^{3}%
\overrightarrow{q}\text{ }\overrightarrow{q}\rho (\overrightarrow{p},%
\overrightarrow{q})
\end{array}
\right.  \label{pqquantummeanvalues}
\end{equation}
where $\rho (\overrightarrow{p},\overrightarrow{q})=<\overrightarrow{p},%
\overrightarrow{q}|\mathbf{D}|\overrightarrow{p},\overrightarrow{q}>$ is a
positive function since $\mathbf{D}$ is a positive operator.

As mentioned in III-B3, we recover that $\rho (\overrightarrow{p},%
\overrightarrow{q})$ is a true classical probability density, but $\hslash $
-dependent, because $\mathbf{D}=\Pi _{S}\mathbf{D}\Pi _{S}$ depends in
general on $\hslash $ through $\Pi _{S}$. Moreover the equations (\ref
{pqquantummeanvalues}) show that the quantum expectation values of $%
\overrightarrow{\mathbf{P}}$ and $\overrightarrow{\mathbf{Q}}$ can always be
expressed with a classical formula, using a $\hslash $ -dependent density of
probability.

More generally, any expectation value $<\mathbf{F}>=Tr(\mathbf{D}.\mathbf{F})
$ of a quantized observable $\mathbf{F}$ obtained from a classical quantity $%
f(\overrightarrow{p},\overrightarrow{q})$ by $\mathbf{F}=\Pi _{S}f(%
\overrightarrow{\mathbf{p}},\overrightarrow{\mathbf{q}})\Pi _{S}$ verifies: 
\begin{equation}
<\mathbf{F}>=\int d^{3}\overrightarrow{p}d^{3}\overrightarrow{q}f(%
\overrightarrow{p},\overrightarrow{q})\rho (\overrightarrow{p},%
\overrightarrow{q})
\end{equation}

So, for any quantum observable $\mathbf{F}$ deducible from a classical one $%
f(\overrightarrow{p},\overrightarrow{q})$ by our procedure of quantization,
the expectation value is obtained by a classical formula, using the density $%
\rho (\overrightarrow{p},\overrightarrow{q})=<\overrightarrow{p},%
\overrightarrow{q}|\mathbf{D}|\overrightarrow{p},\overrightarrow{q}>$.

Nevertheless, these semi-classical expressions cannot be extended to
non-classical observables like $\overrightarrow{\mathbf{S}}$ because we
cannot find any classical quantity associated with $\overrightarrow{\mathbf{S%
}}$.

\section{Quantum Dynamics}

As mentioned in II-B1 and developed in II-B3 and V, Classical Dynamics on $%
\mathcal{H}$ is induced by a classical observable $H$ through the Weak
Dynamical Equations (\ref{meanhamilequs}) that give the evolution of
expectation values. Moreover the evolution of a state $\phi $ is defined by
a unitary operator $U_{t_{1},t_{0}}$ such that $\phi
_{t_{1}}=U_{t_{1},t_{0}}(\phi _{t_{0}})$ and equations (\ref{meanhamilequs})
can be written as equations \ref{newweakdynamics} that we recall here: 
\begin{equation}
\left\{ 
\begin{array}{l}
\frac{d}{dt}<\phi _{t}|\overrightarrow{\mathbf{q}}|\phi _{t}>=<\phi _{t}|%
\overrightarrow{\nabla _{p}}H(\overrightarrow{\mathbf{p}},\overrightarrow{%
\mathbf{q}},t)|\phi _{t}> \\ 
\frac{d}{dt}<\phi _{t}|\overrightarrow{\mathbf{p}}|\phi _{t}>=-<\phi _{t}|%
\overrightarrow{\nabla _{q}}H(\overrightarrow{\mathbf{p}},\overrightarrow{%
\mathbf{q}},t)|\phi _{t}>
\end{array}
\right.  \label{recalldynamics}
\end{equation}

\subsection{The Quantum Operator of Evolution: Schr\"{o}dinger Equation}

Now, since the physical Hilbert space is $\mathcal{H}_{S}$, $\mathcal{H}_{S}$
must be invariant under evolution and then $\Pi
_{S}U_{t_{1},t_{0}}=U_{t_{1},t_{0}}\Pi _{S}$. So for any trajectory of
states $\phi _{t}=U_{t,t_{0}}(\phi _{t_{0}})\in \mathcal{H}_{S}$. Equations (%
\ref{recalldynamics}) of V-A (where quantization is missing) have not to be
changed, but now we must make explicit the condition $|\phi _{t}>=\Pi
_{S}|\phi _{t}>$: 
\begin{equation}
\left\{ 
\begin{array}{l}
\frac{d}{dt}<\phi _{t}|\Pi _{S}\overrightarrow{\mathbf{q}}\Pi _{S}|\phi _{t}>%
\text{ }=\text{ }<\phi _{t}|\Pi _{S}\overrightarrow{\nabla _{p}}H(%
\overrightarrow{\mathbf{p}},\overrightarrow{\mathbf{q}},t)\Pi _{S}|\phi _{t}>
\\ 
\frac{d}{dt}<\phi _{t}|\Pi _{S}\overrightarrow{\mathbf{p}}\Pi _{S}|\phi _{t}>%
\text{ }=-<\phi _{t}|\Pi _{S}\overrightarrow{\nabla _{q}}H(\overrightarrow{%
\mathbf{p}},\overrightarrow{\mathbf{q}},t)\Pi _{S}|\phi _{t}>
\end{array}
\right.  \label{quantumequs1}
\end{equation}

Using the definition of $\overrightarrow{\mathbf{P}}$ and $\overrightarrow{%
\mathbf{Q}}$, equations (\ref{quantumequs1}) become: 
\begin{equation}
\left\{ 
\begin{array}{l}
\frac{d}{dt}<\phi _{t}|\overrightarrow{\mathbf{Q}}|\phi _{t}>\text{ }=\text{ 
}<\phi _{t}|\Pi _{S}\overrightarrow{\nabla _{p}}H(\overrightarrow{\mathbf{p}}
,\overrightarrow{\mathbf{q}},t)\Pi _{S}|\phi _{t}> \\ 
\frac{d}{dt}<\phi _{t}|\overrightarrow{\mathbf{P}}|\phi _{t}>\text{ }=-<\phi
_{t}|\Pi _{S}\overrightarrow{\nabla _{q}}H(\overrightarrow{\mathbf{p}},%
\overrightarrow{\mathbf{q}},t)\Pi _{S}|\phi _{t}>
\end{array}
\right.  \label{quantumequs2}
\end{equation}

Before any further computation, we want first to point out that, by
construction, our new dynamical equations (\ref{quantumequs2}) for
expectation values of $\overrightarrow{\mathbf{P}}$ and $\overrightarrow{%
\mathbf{Q}}$ have always the same semi-classical expression given by (\ref
{meanhamilequs}) where the classical density $\rho _{t}$ is given by $\rho
_{t}(\overrightarrow{p},\overrightarrow{q})=|<\overrightarrow{p},%
\overrightarrow{q}|\phi _{t}>|^{2}$.

Now, in order to write (\ref{quantumequs2}) in a simplified manner, we need
the following technical remarks.\bigskip

\textbf{Remark}: \emph{General Relations between commutators and derivation}

We have seen that any classical quantity $f(\overrightarrow{p},%
\overrightarrow{q})$ is represented on $\mathcal{H}$ by the operator $%
\mathbf{f}=f(\overrightarrow{\mathbf{p}},\overrightarrow{\mathbf{q}})$.
Taking the definition (\ref{quantumgenerators}) of the generators $%
\overrightarrow{\mathbf{P}_{*}}$ and $\overrightarrow{\mathbf{Q}_{*}}$ we
have: 
\begin{equation}
\left\{ 
\begin{array}{l}
\lbrack \overrightarrow{\mathbf{P}_{*}},f(\overrightarrow{\mathbf{p}},%
\overrightarrow{\mathbf{q}})]=-i\hslash \overrightarrow{\nabla _{q}}f(%
\overrightarrow{\mathbf{p}},\overrightarrow{\mathbf{q}}) \\ 
\lbrack \overrightarrow{\mathbf{Q}_{*}},f(\overrightarrow{\mathbf{p}},%
\overrightarrow{\mathbf{q}})]=i\hslash \overrightarrow{\nabla _{p}}f(%
\overrightarrow{\mathbf{\ p}},\overrightarrow{\mathbf{q}})
\end{array}
\right.
\end{equation}

If we project these equations on $\mathcal{H}_{S}$, we obtain: 
\begin{equation}
\left\{ 
\begin{array}{l}
\lbrack \overrightarrow{\mathbf{P}},\Pi _{S}f(\overrightarrow{\mathbf{p}},%
\overrightarrow{\mathbf{q}})\Pi _{S}]=-i\hslash \Pi _{S}\overrightarrow{
\nabla _{q}}f(\overrightarrow{\mathbf{p}},\overrightarrow{\mathbf{q}})\Pi
_{S} \\ 
\lbrack \overrightarrow{\mathbf{Q}},\Pi _{S}f(\overrightarrow{\mathbf{p}},%
\overrightarrow{\mathbf{q}})\Pi _{S}]=i\hslash \Pi _{S}\overrightarrow{
\nabla _{p}}f(\overrightarrow{\mathbf{p}},\overrightarrow{\mathbf{q}})\Pi
_{S}
\end{array}
\right.  \label{relationscommutatorderivation}
\end{equation}

This shows how commutators between $\overrightarrow{\mathbf{P}}$,$%
\overrightarrow{\mathbf{Q}}$ and quantized observables $\Pi _{S}\mathbf{f}%
\Pi _{S}$ are connected to derivations.\bigskip 

We are ready now to write equations (\ref{quantumequs2}) in a simplified
way, using (\ref{relationscommutatorderivation}): 
\begin{equation}
\left\{ 
\begin{array}{l}
\frac{d}{dt}<\phi _{t}|\overrightarrow{\mathbf{Q}}|\phi _{t}>\text{ }
=i\hslash ^{-1}<\phi _{t}|[\Pi _{S}H(\overrightarrow{\mathbf{p}},%
\overrightarrow{\mathbf{q}},t)\Pi _{S},\overrightarrow{\mathbf{Q}}]|\phi
_{t}> \\ 
\frac{d}{dt}<\phi _{t}|\overrightarrow{\mathbf{P}}|\phi _{t}>\text{ }
=i\hslash ^{-1}<\phi _{t}|[\Pi _{S}H(\overrightarrow{\mathbf{p}},%
\overrightarrow{\mathbf{q}},t)\Pi _{S},\overrightarrow{\mathbf{P}}]|\phi
_{t}>
\end{array}
\right.
\end{equation}

So, if we introduce the quantized observable $\mathbf{H}(t)=\Pi _{S}H(%
\overrightarrow{\mathbf{p}},\overrightarrow{\mathbf{q}},t)\Pi _{S}$
associated with the classical Hamiltonian $H(\overrightarrow{p},%
\overrightarrow{q},t)$, previous equations become: 
\begin{equation}
\left\{ 
\begin{array}{l}
\frac{d}{dt}<\phi _{t}|\overrightarrow{\mathbf{Q}}|\phi _{t}>\text{ }%
=i\hslash ^{-1}<\phi _{t}|[\mathbf{H}(t),\overrightarrow{\mathbf{Q}}]|\phi
_{t}> \\ 
\frac{d}{dt}<\phi _{t}|\overrightarrow{\mathbf{P}}|\phi _{t}>\text{ }%
=i\hslash ^{-1}<\phi _{t}|[\mathbf{H}(t),\overrightarrow{\mathbf{P}}]|\phi
_{t}>
\end{array}
\right. 
\end{equation}

Or, for any initial state $\phi _{0}\in \mathcal{H}_{S}$: 
\begin{equation}
\left\{ 
\begin{array}{l}
\frac{d}{dt}<\phi _{0}|U_{t,t_{0}}^{\dagger }\overrightarrow{\mathbf{Q}}%
U_{t,t_{0}}|\phi _{0}>\text{ }=i\hslash ^{-1}<\phi _{0}|U_{t,t_{0}}^{\dagger
}[\mathbf{H}(t),\overrightarrow{\mathbf{Q}}]U_{t,t_{0}}|\phi _{0}> \\ 
\frac{d}{dt}<\phi _{0}|U_{t,t_{0}}^{\dagger }\overrightarrow{\mathbf{P}}%
U_{t,t_{0}}|\phi _{0}>\text{ }=i\hslash ^{-1}<\phi _{0}|U_{t,t_{0}}^{\dagger
}[\mathbf{H}(t),\overrightarrow{\mathbf{P}}]U_{t,t_{0}}|\phi _{0}>
\end{array}
\right. 
\end{equation}

We know that these equations are solved if $U_{t,t_{0}}$ is the unitary
group generated by $\mathbf{H}(t)$: 
\begin{equation}
\left\{ 
\begin{array}{l}
i\hslash \frac{d}{dt}U_{t,t_{0}}=\mathbf{H}(t)U_{t,t_{0}} \\ 
U_{t_{0},t_{0}}=\mathbf{1}
\end{array}
\right.
\end{equation}

So any trajectory of states $|\phi _{t}>$ verifies the equation: 
\begin{equation}
i\hslash \frac{d}{dt}|\phi _{t}>=\mathbf{H}(t)|\phi _{t}>
\label{schrodingerequation}
\end{equation}

This is precisely the general form of Schr\"{o}dinger equation.

So, symbolically speaking, we have solved the problem of Quantum Dynamics.

To finish, let us remark that the expectation value $<\phi _{t}|\mathbf{H}%
(t)|\phi _{t}>$ possesses also a semi-classical expression: 
\begin{equation}
<\phi _{t}|\mathbf{H}(t)|\phi _{t}>=\int d^{3}\overrightarrow{p}d^{3}%
\overrightarrow{q}\rho _{t}\text{ }H(\overrightarrow{p},\overrightarrow{q},t)
\end{equation}
where $\rho _{t}(\overrightarrow{p},\overrightarrow{q})=|<\overrightarrow{p},%
\overrightarrow{q}|\phi _{t}>|^{2}$, because $\mathbf{H}(t)=\Pi _{S}H(%
\overrightarrow{\mathbf{p}},\overrightarrow{\mathbf{q}},t)\Pi _{S}$ is the
quantized version of the classical Hamiltonian (cf. X-E).

\textbf{Conclusion}\newline
Of course we have recovered that the evolution of states is given by
Schr\"{o}dinger equation. But we want to remark that our starting point for
Dynamics was uniquely the ''weak dynamical equations'' introduced early in
II-B3, and by construction they are always valid using the pseudo-classical
density $\rho _{t}(\overrightarrow{p},\overrightarrow{q})=|<\overrightarrow{p%
},\overrightarrow{q}|\phi _{t}>|^{2}$. So, we have shown in fact, that weak
dynamical equations possess solutions in $\rho _{t}(\overrightarrow{p},%
\overrightarrow{q})$ that do not follow Liouville equation, and these other
solutions are those given by Quantum Dynamics.

Of course, it remains to make explicit the quantum Hamiltonian $\mathbf{H}
(t) $ to prove that we recover usual expression of Schr\"{o}dinger equation.
This is done in the last paragraph.

\subsection{The Quantum Hamiltonian}

We have recalled in II-B1 that the general form of the classical Hamiltonian 
$H(\overrightarrow{p},\overrightarrow{q},t)$ is $H=\frac{1}{2M}\left( 
\overrightarrow{p}-e\overrightarrow{A}(\overrightarrow{q},t)\right) ^{2}+V(%
\overrightarrow{q},t)$, the quantum Hamiltonian being $\mathbf{H}(t)=\Pi
_{S}H(\overrightarrow{\mathbf{p}},\overrightarrow{\mathbf{q}},t)\Pi _{S}$.

To simplify computations, we study independently the free Hamiltonian, the
case of interaction with a potential energy and the case of interaction with
a magnetic field.

\subsubsection{The free Quantum Hamiltonian}

The quantum hamiltonian $\mathbf{H}$ reduces to: 
\begin{equation}
\mathbf{H}=\frac{1}{2M}\Pi _{S}\overrightarrow{\mathbf{p}}^{2}\Pi _{S}
\end{equation}

Taking into account the expression of $<\overrightarrow{p},\overrightarrow{q}
|\overrightarrow{x},S,m>$, we compute the matrix element $<\overrightarrow{x}
,S,m|\overrightarrow{\mathbf{p}}^{2}|\overrightarrow{y},S,m_{1}>$ and we
find: 
\begin{equation}
<\overrightarrow{x},S,m|\overrightarrow{\mathbf{p}}^{2}|\overrightarrow{y}
,S,m_{1}>=-\hslash ^{2}(\Delta \delta )(\overrightarrow{x}-\overrightarrow{y}
)h^{3}\int d^{3}\overrightarrow{q}\Phi _{S,m}^{*}(\overrightarrow{x}-%
\overrightarrow{q})\Phi _{S,m_{1}}(\overrightarrow{y}-\overrightarrow{q})
\end{equation}

From the properties of $\Delta \delta $, this expression can be splitted
into: 
\begin{equation}
\left\{ 
\begin{array}{l}
<\overrightarrow{x},S,m|\overrightarrow{\mathbf{p}}^{2}|\overrightarrow{y}
,S,m_{1}>=A+B+C \\ 
A=-\hslash ^{2}(\Delta \delta )(\overrightarrow{x}-\overrightarrow{y}
)h^{3}\int d^{3}\overrightarrow{q}\Phi _{S,m}^{*}(\overrightarrow{x}-%
\overrightarrow{q})\Phi _{S,m_{1}}(\overrightarrow{x}-\overrightarrow{q}) \\ 
B=-2\hslash ^{2}(\overrightarrow{\nabla }\delta )(\overrightarrow{x}-%
\overrightarrow{y})h^{3}\int d^{3}\overrightarrow{q}\Phi _{S,m}^{*}(%
\overrightarrow{x}-\overrightarrow{q})(\overrightarrow{\nabla }\Phi
_{S,m_{1}})(\overrightarrow{x}-\overrightarrow{q}) \\ 
C=-\hslash ^{2}\delta (\overrightarrow{x}-\overrightarrow{y})h^{3}\int d^{3}%
\overrightarrow{q}\Phi _{S,m}^{*}(\overrightarrow{x}-\overrightarrow{q}
)(\Delta \Phi _{S,m_{1}})(\overrightarrow{x}-\overrightarrow{q})
\end{array}
\right.
\end{equation}

Taking into account the properties of orthogonality and normalization of the
fields $\Phi _{S,m}$, we obtain first: 
\begin{equation}
A=-\hslash ^{2}(\Delta \delta )(\overrightarrow{x}-\overrightarrow{y})\delta
_{m,m_{1}}
\end{equation}

Now, because of the parity of $\Phi _{S,m}$, we have: 
\begin{equation}
B=0
\end{equation}

Finally, if we use the explicit expression (\ref{irreduciblestates}) of $%
\Phi _{S,m}$ and the expression of $\Delta $ in spherical coordinates, we
find: 
\begin{equation}
\left\{ 
\begin{array}{l}
C=K\delta (\overrightarrow{x}-\overrightarrow{y})\delta _{m,m_{1}} \\ 
K=\hslash ^{2}h^{3}\int_{0}^{\infty }dr\left\{ [(r\Psi )^{\prime
}]^{2}+S(S+1)\Psi ^{2}\right\}
\end{array}
\right.
\end{equation}

Now, we want to use the scaling on the function $\Psi (x)$ introduced in IX
to exhibit the dependence of the representation in a scale length $\lambda $
. We recall that we take an undimensional mathematical function $\Psi _{0}$
such that $\Psi (x)=(\lambda h)^{-3/2}\Psi _{0}(x/\lambda )$. Using this
scaling, we find that the previous constant $K$ is: 
\begin{equation}
\left\{ 
\begin{array}{l}
K=(\hslash \chi /\lambda )^{2} \\ 
\chi ^{2}=\int_{0}^{\infty }du\left\{ [(u\Psi _{0})^{\prime
}]^{2}+S(S+1)\Psi _{0}^{2}\right\}
\end{array}
\right.
\end{equation}
where $\chi ^{2}$ is an undimensional positive coefficient.

If we summarize this computation, we have: 
\begin{equation}
\left\{ 
\begin{array}{l}
<\overrightarrow{x},S,m|\mathbf{H}|\overrightarrow{y},S,m_{1}>=-(\hslash
^{2}/2M)(\Delta \delta )(\overrightarrow{x}-\overrightarrow{y})\delta
_{m,m_{1}}+E_{0}\delta (\overrightarrow{x}-\overrightarrow{y})\delta
_{m,m_{1}} \\ 
E_{0}=\hslash ^{2}\chi ^{2}\lambda ^{-2}/2M
\end{array}
\right.
\end{equation}

Or in the operator formalism: 
\begin{equation}
\mathbf{H=}\frac{1}{2M}\overrightarrow{\mathbf{P}}^{2}+E_{0}
\end{equation}

We recover the usual quantum Hamiltonian of the free particle with a
supplementary constant $E_{0}$. Of course this constant does not modify
dynamical properties and we can simply ignore it.

But if we want to give a physical meaning to this term, we must say that the
particle possesses an intrinsic proper energy. So, exactly as in paragraph
IX, we must distinguish the case of true elementary particles from the case
of more complex systems.

\textbf{The case of elementary particles}

For a true particle, the only possible definition of $E_{0}$ is the mass
energy of the particle. Of course, we cannot directly find $E_{0}$ in the
frame of non-relativistic mechanics and then $E_{0}=Mc^{2}$ must be induced
from external arguments. But if we assume this formula, we find that the
scale length $\lambda $ verifies: 
\begin{equation}
\lambda =\lambda _{c}\frac{\chi }{2\sqrt{2}\pi }
\end{equation}
where $\lambda _{c}=h/Mc$ is the Compton wave length.

So we recover the intuitive result of paragraph IX, where we have recognized
in $\lambda $ the Compton wave length. Since we are in non-relativistic
mechanics, $\lambda $ must always be a very small quantity in regard to all
other length scales.

\textbf{The case of complex systems}

As mentioned in IX, if the ''particle'' possesses an internal structure, we
must assume that all internal degrees of freedom are frozen, and then $E_{0}$
represents the internal energy.

\subsubsection{Quantum Hamiltonian with a potential energy}

The quantum Hamiltonian $\mathbf{H}(t)$ reduces to: 
\begin{equation}
\mathbf{H}(t)=\frac{1}{2M}\Pi _{S}\overrightarrow{\mathbf{p}}^{2}\Pi
_{S}+\Pi _{S}V(\overrightarrow{\mathbf{q}},t)\Pi _{S}
\end{equation}

Taking into account the result on the free case, we have: 
\begin{equation}
\mathbf{H}(t)=\frac{1}{2M}\overrightarrow{\mathbf{P}}^{2}+E_{0}+\Pi _{S}V(%
\overrightarrow{\mathbf{q}},t)\Pi _{S}
\end{equation}

So we have only to specify the matrix element $<\overrightarrow{x},S,m|V(%
\overrightarrow{\mathbf{q}},t)|\overrightarrow{y},S,m_{1}>$.

After a few algebra, we find: 
\begin{equation}
<\overrightarrow{x},S,m|V(\overrightarrow{\mathbf{q}},t)|\overrightarrow{y}
,S,m_{1}>=\delta (\overrightarrow{x}-\overrightarrow{y})h^{3}\int d^{3}%
\overrightarrow{q}V(\overrightarrow{x}-\overrightarrow{q},t)\Phi _{S,m}^{*}(%
\overrightarrow{q})\Phi _{S,m_{1}}(\overrightarrow{q})
\end{equation}

Using our scaling on the function $\Psi $ in $\Phi _{S,m}$, we have: 
\begin{equation}
<\overrightarrow{x},S,m|V(\overrightarrow{\mathbf{q}},t)|\overrightarrow{y}
,S,m_{1}>=\delta (\overrightarrow{x}-\overrightarrow{y})\int d^{3}%
\overrightarrow{u}V(\overrightarrow{x}-\lambda \overrightarrow{u},t)\Psi
_{0}(u)^{2}Y_{S}^{m}(\hat{u})^{*}Y_{S}^{m_{1}}(\hat{u})  \label{equv(q)quant}
\end{equation}
where $\hat{u}=\overrightarrow{u}/|\overrightarrow{u}|$.

Now, if we want to simplify the previous expression, we must take into
account the magnitude of $\lambda $ in comparison with the length scale of
variation of $V(\overrightarrow{x})$. We have seen in the previous paragraph
that $\lambda $ must always be very small in regard to all other length
scales, so we can use the following development in the equation (\ref
{equv(q)quant} ): 
\begin{equation}
V(\overrightarrow{x}-\lambda \overrightarrow{u},t)\simeq V(\overrightarrow{x}
,t)-\lambda \overrightarrow{u}.\overrightarrow{\nabla }V(\overrightarrow{x}
,t)+\lambda ^{2}\epsilon
\end{equation}

Taking into account the parity of the spherical harmonics, we obtain that
the first order in $\lambda $ vanishes and: 
\begin{equation}
<\overrightarrow{x},S,m|V(\overrightarrow{\mathbf{q}},t)|\overrightarrow{y}
,S,m_{1}>=V(\overrightarrow{x},t)\delta (\overrightarrow{x}-\overrightarrow{y%
})\delta _{m,m_{1}}+\lambda ^{2}\epsilon
\end{equation}

Then, up to the second order in $\lambda $, the operator $\Pi _{S}V(%
\overrightarrow{\mathbf{q}},t)\Pi _{S}$ can be identified with $V(%
\overrightarrow{\mathbf{Q}},t)$.

So if we neglect the corrections in $\lambda ^{2}$, we conclude that the
quantum Hamiltonian $\mathbf{H}(t)$ is: 
\begin{equation}
\mathbf{H}(t)=\frac{1}{2M}\overrightarrow{\mathbf{P}}^{2}+E_{0}+V(%
\overrightarrow{\mathbf{Q}},t)
\end{equation}

We recover the usual Hamiltonian of Quantum Mechanics.

\textbf{The Harmonic case}\newline
In the particular case where $V(\overrightarrow{\mathbf{q}})=\frac{1}{2}
m\omega ^{2}\overrightarrow{\mathbf{q}}^{2}$, we can compute completely the
operator $\Pi _{S}V(\overrightarrow{\mathbf{q}})\Pi _{S}$ and we obtain: 
\begin{equation}
\Pi _{S}V(\overrightarrow{\mathbf{q}})\Pi _{S}=V(\overrightarrow{\mathbf{Q}}
)+E_{1}
\end{equation}
where $E_{1}$ is a supplementary constant energy given by: 
\begin{equation}
E_{1}=\frac{1}{2}m\omega ^{2}\lambda ^{2}\eta ^{2}\text{ with }\eta
^{2}=\int_{0}^{\infty }u^{4}du\Psi _{0}(u)^{2}
\end{equation}

\textbf{Conclusion}\newline
We recover the quantum operator of potential energy, but we see that the
operator $V(\overrightarrow{\mathbf{Q}},t)$ is only an approximation (up to
second order in $\lambda $) of the true quantized operator $\Pi _{S}V(%
\overrightarrow{\mathbf{q}},t)\Pi _{S}$. So, the prescription of the
correspondence principle corresponds in fact to the limit case $\lambda
\rightarrow 0$. But, if it is possible to take this limit for $\Pi _{S}V(%
\overrightarrow{\mathbf{q}},t)\Pi _{S}$, we cannot do that directly with the
free part of the Hamiltonian because $E_{0}\varpropto 1/\lambda ^{2}$: we
need first to renormalize the free Hamiltonian. Moreover, if we take this
limit, we loose the ''connection with classical world'' because the value $%
\lambda =0$ is forbidden for reasons of normalization: $\lambda $ can be as
small as you want but never cancelled.\newline
To conclude, we can say that this approach specifies the limit of the usual
''correspondence principle'' that postulates that the classical potential $V(%
\overrightarrow{q},t)$ must be directly lifted into quantum Hamiltonian. In
fact, mathematically speaking, the true operator $\Pi _{S}V(\overrightarrow{%
\mathbf{q}},t)\Pi _{S}$ depends on $S$ and $\lambda $, but because $\lambda $
is always very small (in non-relativistic mechanics), we can physically
ignore it.

\subsubsection{Quantum hamiltonian with a magnetic field}

To simplify computations, we only study the case of a uniform magnetic field 
$\overrightarrow{B}$ associated with the potential vector $\overrightarrow{A}
(\overrightarrow{q})=\frac{1}{2}\overrightarrow{B}\wedge \overrightarrow{q}$
. The quantum Hamiltonian $\mathbf{H}$ is: 
\begin{equation}
\mathbf{H}=\frac{1}{2M}\Pi _{S}[\overrightarrow{\mathbf{p}}-e\overrightarrow{
A}(\overrightarrow{\mathbf{q}})]^{2}\Pi _{S}
\end{equation}

Developing the previous expression, we have: 
\begin{equation}
\mathbf{H}=\frac{1}{2M}\Pi _{S}\overrightarrow{\mathbf{p}}^{2}\Pi _{S}-\frac{
e}{M}\Pi _{S}\overrightarrow{\mathbf{p}}.\overrightarrow{A}(\overrightarrow{%
\mathbf{q}})\Pi _{S}+\frac{e^{2}}{2M}\Pi _{S}\overrightarrow{A}(%
\overrightarrow{\mathbf{q}})^{2}\Pi _{S}
\end{equation}

Using the expression of the free Hamiltonian we obtain: 
\begin{equation}
\left\{ 
\begin{array}{l}
\mathbf{H}=\frac{1}{2M}\overrightarrow{\mathbf{P}}^{2}+E_{0}+\mathbf{H}_{1}+%
\mathbf{H}_{2} \\ 
\mathbf{H}_{1}=-\frac{e}{M}\Pi _{S}\overrightarrow{\mathbf{p}}.%
\overrightarrow{A}(\overrightarrow{\mathbf{q}})\Pi _{S} \\ 
\mathbf{H}_{2}=\frac{e^{2}}{2M}\Pi _{S}\overrightarrow{A}(\overrightarrow{%
\mathbf{q}})^{2}\Pi _{S}
\end{array}
\right.
\end{equation}

We first look at $\mathbf{H}_{1}$, using the explicit form of $%
\overrightarrow{A}(\overrightarrow{q})$: 
\begin{equation}
\mathbf{H}_{1}=-\frac{e}{2M}\Pi _{S}\overrightarrow{\mathbf{p}}.(%
\overrightarrow{B}\wedge \overrightarrow{\mathbf{q}})\Pi _{S}=-\frac{e}{2M}%
\overrightarrow{B}.\Pi _{S}\overrightarrow{\mathbf{q}}\wedge \overrightarrow{%
\mathbf{p}}\Pi _{S}
\end{equation}

But we have seen in paragraph X-B that $\Pi _{S}\overrightarrow{\mathbf{q}}
\wedge \overrightarrow{\mathbf{p}}\Pi _{S}=\overrightarrow{\mathbf{Q}}\wedge 
\overrightarrow{\mathbf{P}}-\overrightarrow{\mathbf{S}}$, then: 
\begin{equation}
\mathbf{H}_{1}=-\frac{e}{2M}\overrightarrow{B}.(\overrightarrow{\mathbf{Q}}
\wedge \overrightarrow{\mathbf{P}})+\frac{e}{2M}\overrightarrow{B}.%
\overrightarrow{\mathbf{S}}
\end{equation}

And we can transform again the first term to obtain: 
\begin{equation}
\mathbf{H}_{1}=-\frac{e}{M}(\overrightarrow{A}(\overrightarrow{\mathbf{Q}}).%
\overrightarrow{\mathbf{P}}+\overrightarrow{\mathbf{P}}.\overrightarrow{A}(%
\overrightarrow{\mathbf{Q}})+\frac{e}{2M}\overrightarrow{B}.\overrightarrow{%
\mathbf{S}}
\end{equation}

Since $\mathbf{H}_{2}$ can be seen as an harmonic potential energy, we can
use the result of the previous paragraph: 
\begin{equation}
\mathbf{H}_{2}=\frac{e^{2}}{2M}\overrightarrow{A}(\overrightarrow{\mathbf{Q}}
)^{2}+E_{1}
\end{equation}

If we collect all the results, we conclude that the quantum hamiltonian is: 
\begin{equation}
\mathbf{H=}\frac{1}{2M}[\overrightarrow{\mathbf{P}}-e\overrightarrow{A}(%
\overrightarrow{\mathbf{Q}})]^{2}+\frac{e}{2M}\overrightarrow{B}.%
\overrightarrow{\mathbf{S}}+E_{0}+E_{1}
\end{equation}

So we find that our projection of the classical hamiltonian (where spin is
missing) generates directly an interaction between spin and magnetic field.
Of course, if we believe in this formula, we must say that the particle
possesses a magnetic momentum $\overrightarrow{\mathbf{\mu }_{0}}=-\frac{e}{
2M}\overrightarrow{\mathbf{S}}$. Unfortunately $\overrightarrow{\mathbf{\mu }
_{0}}$ does not correspond in general to the true value of $\overrightarrow{%
\mathbf{\mu }}$. The reason is that $\overrightarrow{\mathbf{\mu }_{0}}$
only describes the part of $\overrightarrow{\mathbf{\mu }}$ deducible from
our classical Hamiltonian.

In fact, when $\overrightarrow{B}$ is uniform, we can add to $\mathbf{H}$
any supplementary term as $\mathbf{H}_{I}=-(g+1)\frac{e}{2M}\overrightarrow{B%
}.\overrightarrow{\mathbf{S}}$ without changing the equations of motion for $%
\overrightarrow{\mathbf{P}}$ and $\overrightarrow{\mathbf{Q}}$. So, because $%
\overrightarrow{\mathbf{S}}$ is not a classical observable, the part of $%
\mathbf{H}$ that specifies the evolution of $\overrightarrow{\mathbf{S}}$ is
not given by our procedure which is only based on the classical observables $%
\overrightarrow{\mathbf{P}}$ and $\overrightarrow{\mathbf{Q}}$. The term $-%
\overrightarrow{\mathbf{\mu }_{0}}.\overrightarrow{B}$ must be only seen as
an indication on the form of interaction between $\overrightarrow{\mathbf{S}}
$ and $\overrightarrow{B}$. Of course, adding $\mathbf{H}_{I}$ to our
Hamiltonian allows to recover the usual formula for $\overrightarrow{\mathbf{%
\mu }}$.

Finally, if we assume $\overrightarrow{B}$ to be non-uniform, we see that
the new Hamiltonian $\mathbf{H}_{I}$ introduces a new dynamical coupling
between classical observables and $\overrightarrow{\mathbf{S}}$. This effect
(on expectation values) is purely quantic and cannot be reproduced by our
procedure.

\section{Conclusion}

This analysis shows how the overlapped components of Classical and Quantum
Mechanics can be separated to give a complete sequential structure that
allows a better understanding of the role of each ingredient. Moreover, we
can get rid of too crude rules of quantization based only on pairs of
canonical coordinates, that do not explain why only one system of canonical
coordinates gives the right quantization. Our procedure allows also to give
a satisfactory explanation of a apparent illogical process that consists in
building quantum dynamical equations only using classical quantities (the
classical potential energy for example). Moreover, we recover the central
part played by coherent states to connect classical and quantum objects.

Nevertheless, as indicated in the introduction, this article is not
''logically complete'' because it should be more enlightning to recover
first general quantum axiomatics (as a change of mathematical language) from
ideas of Classical Mechanics. So, we can look at the beginnning of this
article as a ''middle-point'', the first part will be published later.

To conclude, we can consider two natural directions of generalizations of
our procedure: the first one concerns the multiparticle case and second
quantization, the second one is of course Special Relativity.\bigskip 
\newline
{\LARGE Acknowledgement}\newline
It is a pleasure to acknowledge useful discussions with Dr. A. Valance and
Dr. J. Mourad.

\end{document}